\documentclass[sort&compress,final,3p,times,twocolumn,fixfloat]{elsarticle}

\usepackage{graphicx}
\usepackage{amssymb} 
\usepackage{amsmath} 
\usepackage{times}
\usepackage{tabularx}

\usepackage{scalerel}
\usepackage{tikz}
\usetikzlibrary{svg.path}

\definecolor{orcidlogocol}{HTML}{A6CE39}
\tikzset{
  orcidlogo/.pic={
    \fill[orcidlogocol] svg{M256,128c0,70.7-57.3,128-128,128C57.3,256,0,198.7,0,128C0,57.3,57.3,0,128,0C198.7,0,256,57.3,256,128z};
    \fill[white] svg{M86.3,186.2H70.9V79.1h15.4v48.4V186.2z}
                 svg{M108.9,79.1h41.6c39.6,0,57,28.3,57,53.6c0,27.5-21.5,53.6-56.8,53.6h-41.8V79.1z M124.3,172.4h24.5c34.9,0,42.9-26.5,42.9-39.7c0-21.5-13.7-39.7-43.7-39.7h-23.7V172.4z}
                 svg{M88.7,56.8c0,5.5-4.5,10.1-10.1,10.1c-5.6,0-10.1-4.6-10.1-10.1c0-5.6,4.5-10.1,10.1-10.1C84.2,46.7,88.7,51.3,88.7,56.8z};
  }
}

\newcommand\orcidicon[1]{\href{https://orcid.org/#1}{\mbox{\scalerel*{
\begin{tikzpicture}[yscale=-1,transform shape]
\pic{orcidlogo};
\end{tikzpicture}
}{|}}}}

\usepackage{txfonts}
\usepackage{lineno}

\journal{Physics Letters B}

 \usepackage{ifpdf}
 \ifpdf
 \usepackage[pdftex]{hyperref}
 \else
 \usepackage[hypertex]{hyperref}
 \fi

 \hypersetup{
   pdftitle={},%
   pdfauthor={},%
   pdfsubject={},%
   pdfkeywords={},%
   pdfstartview={},%
   bookmarksopen=true, breaklinks=true, debug=true, %
   colorlinks=true, linkcolor=red, citecolor=blue, urlcolor=blue
 }

\usepackage{float} 
\usepackage[caption=false]{subfig} 

\newcount\savefnused
\newcount\savefndone

\newcommand{\savefootnote}[2][\empty]
{\ifx\empty#1\footnotemark\else\footnotemark[#1]\fi
 \global\advance\savefnused by 1
 \expandafter\xdef\csname savefnmark\the\savefnused\endcsname{\thefootnote}%
 \expandafter\xdef\csname savefntext\the\savefnused\endcsname{#2}%
}
\newcommand{\flushfootnote}{\loop\ifnum\savefndone<\savefnused
  \global\advance\savefndone by 1
  \footnotetext[\csname savefnmark\the\savefndone\endcsname]%
    {\csname savefntext\the\savefndone\endcsname}%
  \global\expandafter\let\csname savefnmark\the\savefndone\endcsname\relax
  \global\expandafter\let\csname savefntext\the\savefndone\endcsname\relax
\repeat}

\usepackage{multirow,bigdelim}

\usepackage{textcomp}
\usepackage{comment}

\usepackage{lineno}

\usepackage{float} 
\usepackage{subfig} 
\usepackage{amsmath} %
\usepackage{amstext}

\usepackage{graphicx}%
\usepackage{tabularx,ragged2e}%

\newcolumntype{Y}{>{\centering\arraybackslash}X}

\newcommand{\eqs}[1]{\begin{equation} \begin{split} #1\end{split} \end{equation} }

\def\ie{{\it i.e.}}
\def\eg{{\it e.g.}}

\def\GeV{{\rm GeV}}
\def\GeV2{{\rm GeV}^2}

\newcommand{\Q}{{\cal Q}}
\renewcommand{\P}{{\cal P}}

\newcommand{\ce}[1]{Eq.~(\ref{#1})}
\newcommand{\cf}[1]{{Fig.~\ref{#1}}}
\newcommand{\ct}[1]{{Table~\ref{#1}}}

\begin{document} 
\begin{frontmatter}

\title{
Complete NLO QCD study of single- and double-quarkonium hadroproduction in the colour-evaporation model at the Tevatron and the LHC
}
\author[a]{Jean-Philippe Lansberg\orcidicon{0000-0003-2746-5986}}
\author[b]{Hua-Sheng Shao\orcidicon{0000-0002-4158-0668}}
\author[c,a,d]{Nodoka Yamanaka\orcidicon{0000-0002-0602-8288}}
\author[e,f]{Yu-Jie Zhang\orcidicon{0000-0002-2748-3300}}
\author[g]{Camille No\^us\orcidicon{0000-0002-0778-8115}}
\address[a]{Universit\'e Paris-Saclay, CNRS, IJCLab, 91405 Orsay, France}
\address[b]{Laboratoire de Physique Th\'eorique et Hautes Energies (LPTHE), UMR 7589, Sorbonne Universit\'e et CNRS, 4 place Jussieu, 75252 Paris, France}
\address[c]{Amherst Center for Fundamental Interactions, Department of Physics,
University of Massachusetts Amherst, MA 01003, USA}
\address[d]{Theoretical Research Division, Nishina Center, RIKEN, Saitama 351-0198, Japan}
\address[e]{School of Physics and Nuclear Energy Engineering, Beihang University, Beijing 100083, China}
\address[f]{Center for High Energy Physics, Peking University, Beijing 100871, China}
\address[g]{Cogitamus Laboratory}
\begin{abstract}
{\small
We study the Single-Parton-Scattering (SPS) production of double quarkonia ($J/\psi+J/\psi$, $J/\psi+\Upsilon$, and $\Upsilon+\Upsilon$) in $pp$ and $p \bar p$ collisions at the LHC and the Tevatron as measured by the CMS, ATLAS, LHCb, and D0 experiments
in the Colour-Evaporation Model (CEM), based on the quark-hadron-duality, including Next-to-Leading Order (NLO) 
QCD corrections up to $\alpha_s^5$.
To do so, we also perform the first true NLO --up to $\alpha_s^4$-- study of the $p_T$-differential cross section for
single-quarkonium production. This allows us to fix the non-perturbative CEM parameters at NLO accuracy in the region
where quarkonium-pair data are measured. Our results show that the CEM at NLO in general significantly undershoots 
these experimental data and, in view of the other existing SPS studies, confirm the need for Double Parton Scattering (DPS) to account for the data.
Our NLO study of single-quarkonium production at mid and large $p_T$ also confirms the difficulty of the approach to 
account for the measured $p_T$ spectra; this is reminiscent of the impossibility to fit single-quarkonium data with the sole
$^3S_1^{[8]}$ NRQCD contribution from gluon fragmentation. We stress that the discrepancy occurs in a kinematical region where the new features of the improved CEM are not relevant.
}
\end{abstract}

\date{\today}
\end{frontmatter}
%

\section{Introduction}

Quarkonium-pair production in high-energy hadron-hadron collisions is an interesting probe of many physics
phenomena. It can help us study the physics underlying double parton scatterings (DPS)~\cite{Kom:2011bd,Lansberg:2014swa}, thence the gluon-gluon correlations in the proton (see \eg~\cite{Blok:2011bu,Ryskin:2012qx,Gaunt:2012dd,Blok:2013bpa,Diehl:2014vaa,Blok:2016lmd,Diehl:2017wew,Kasemets:2017vyh,Rinaldi:2018slz,Rinaldi:2018bsf,Nagar:2019njl,Cotogno:2020iio}). It can also provide us with unique 
information about the distribution of linearly-polarised gluons inside the proton~\cite{Lansberg:2017dzg,Scarpa:2019ucf}. 
Finally, it remains a crucial test of quarkonium-production models~(see~\cite{Lansberg:2006dh,Brambilla:2010cs,Andronic:2015wma,Lansberg:2019adr} for reviews) which should of course account both for single- 
and double-quarkonium yields as well as associated production~\cite{Lansberg:2019adr}. Going further, triple-$J/\psi$ production should also help us probe both DPS and Triple Parton Scatterings (TPS)~\cite{dEnterria:2016ids,Shao:2019qob}.

In the recent years, there has been an accumulation of experimental hints~\cite{Aaij:2011yc,Abazov:2014qba, Khachatryan:2014iia,Lansberg:2014swa,Abazov:2015fbl,Shao:2016wor,Aaboud:2016fzt,Aaij:2016bqq,Sirunyan:2020txn} that quarkonium pairs
can be produced in a significant amount by two simultaneous parton-parton scatterings -- the DPS. This is particularly true 
at large rapidity separations, $\Delta y_{\psi\psi}$,  where the {\it a priori} leading Single Parton Scatterings (SPS) are suppressed
since they generate highly correlated quarkonium pairs, thus at low $\Delta y_{\psi\psi}$. The region of large $\Delta y_{\psi\psi}$ is therefore a good candidate for a control region for DPS extraction.

Di-quarkonia from DPS are in principle fully decorrelated. Such a property was in fact used to disentangle their contributions
from those of the SPS. For instance, one expects a flat event distribution as a function of the azimuthal angle between 
both quarkonia, $\Delta \phi_{\psi\psi}$. However, a reliable SPS-DPS separation often calls for a good control of the SPS kinematical distribution which
can be similar to that of the DPS in some phase-space regions. Many theoretical di-quarkonium SPS studies have been carried out~\cite{Kartvelishvili:1984ur,Humpert:1983yj,Vogt:1995tf,Li:2009ug,Qiao:2009kg,Ko:2010xy,Baranov:2011zz,Baranov:2012re,Berezhnoy:2011xy,Li:2013csa,Lansberg:2013qka,%
Lansberg:2014swa,Sun:2014gca,He:2015qya,Baranov:2015cle,Lansberg:2015lva,Likhoded:2016zmk,Lansberg:2017dzg,%
Cisek:2017ikn,He:2018hwb,He:2019qqr,Babiarz:2019pth} but only a few~\cite{Baranov:2012re,Lansberg:2013qka,Sun:2014gca,Lansberg:2014swa,%
Likhoded:2016zmk,Shao:2016wor,Lansberg:2019fgm} dealt with  QCD corrections, some
of which might be relevant where the DPS yields are found to be large.  

At small $\Delta y_{\psi\psi}$, all the experimental data sets are in fact in good agreement with the SPS predictions 
from the Colour-Singlet Model (CSM), \ie\ the LO in the heavy-quark relative velocity, $v$, expansion of Non Relativistic QCD 
(NRQCD)~\cite{Bodwin:1994jh}. These predictions are known up to NLO accuracy~\cite{Lansberg:2013qka,Sun:2014gca,Likhoded:2016zmk}.
In addition, the NRQCD Colour-Octet (CO) contributions are found to be negligible\footnote{This remains true~\cite{Lansberg:2019fgm} whatever  values of the NRQCD Long Distance Matrix Elements (LDMEs) are used -- provided of course that they account for the majority of the corresponding existing single-quarkonium production data.} in this region~\cite{Lansberg:2014swa,He:2015qya,Lansberg:2019fgm} which is in line with the expected suppression by $\mathcal{O}(v^8)$ with respect to the CS contributions.

For increasing $\Delta y_{\psi\psi}$, the lack of complete NLO NRQCD studies is prejudicial and opens the door to some debates~\cite{Lansberg:2014swa,He:2015qya,Lansberg:2019fgm} about the possibility for unexpectedly large SPS contributions from CO contributions. Indeed, owing to the large uncertainties in the LDME determinations~\cite{Lansberg:2019fgm}, NRQCD at LO shows a very low predictive power, \eg\ in regions where the DPS is thought to be the dominant source of quarkonium pairs. Hopefully, possible future NLO studies could close this debate.

This is important since recent direct and indirect DPS extractions based on quarkonia in pairs~\cite{Lansberg:2014swa,Abazov:2015fbl,Shao:2016wor,Aaboud:2016fzt} or in association with a vector boson~\cite{Lansberg:2016rcx,Lansberg:2017chq}
seem to point~\cite{Lansberg:2017chq} at an unexpected flavour or momentum dependence of the parton correlations in the proton --as encoded in the well known quantity $\sigma_{\rm eff.}$~\cite{Rinaldi:2018slz}--,
when compared to other (direct and indirect) extractions~\cite{Akesson:1986iv,Alitti:1991rd,Abe:1993rv,Abe:1997xk,Abazov:2009gc,Aaij:2012dz,Aad:2013bjm,Chatrchyan:2013xxa,Aaij:2015wpa,Aaboud:2016dea,%
Lansberg:2016muq,Sirunyan:2017hlu}.

Using the colour-evaporation model (CEM)~\cite{Fritzsch:1977ay,Halzen:1977rs} --a model based on quark-hadron duality but which shares some features of NRQCD~\cite{Bodwin:2005hm}, in particular the direct production of vector quarkonia from gluon fragmentation-- we wish to advance our understanding of the SPS contributions to quarkonium pairs. 
In addition, its implementation is very similar to that of open heavy-flavour production
and can be done in {\small \sc MadGraph5\_aMC@NLO}~\cite{Alwall:2014hca} with some tunings. Finally, 
it is straightforward to treat the feed-down contributions (\eg\ from $\chi_c$) to prompt $J/\psi$ in the CEM.
Altogether, this allows us to perform the first complete NLO study of quarkonium-pair production using one of the widely used 
quarkonium-production models. 

In the case of $J/\psi+Z$~\cite{Lansberg:2016rcx} and $J/\psi+W$~\cite{Lansberg:2017chq} production, 
we have shown that the CEM provides an upper limit on the SPS contributions. This is also likely the case for $J/\psi+\Upsilon$
production and for the $J/\psi+J/\psi$ case in the kinematical region where gluon fragmentation to both quarkonia is expected to be
dominant. More generally, it offers an indirect way to scrutinise whether some specific configurations of the 
heavy-quark pair receive at NLO kinematically-enhanced contributions, which result in large $K$ factors (see \eg~\cite{Lansberg:2013qka}). Indeed, if we were to observe a large $K$ factor to the di-quarkonium CEM yields, where all the pair (spin and colour) configurations are summed with the same weights, this
would necessarily signal a potential large $K$ factor to some NRQCD contributions.

Such a complete NLO study for di-quarkonium necessitates a coherent determination of the non-perturbative CEM parameters -- one per particle. Therefore, an interesting side product of our study is the corresponding  NLO study of the 
$p_T$-differential cross section of {\it single} $J/\psi$, $\psi(2S)$ and $\Upsilon(nS)$ hadroproduction. 
To what concerns the $\psi(2S)$ and $\Upsilon(nS)$, this is the very first study of this kind. 
So far the  NLO CEM studies~\cite{Bedjidian:2004gd,Nelson:2012bc} were held for the $p_T$-integrated yield at $\alpha_s^3$.

This paper is organised as follows.
In section \ref{sec:NLOCEM}, we explain the methodology of our NLO CEM calculation. In section~\ref{sec:fitCEM},
we discuss our original results for the $p_T$ distribution of single $J/\psi$, $\psi(2S)$ and $\Upsilon(nS)$
in the CEM at NLO, which we use to fit the corresponding non-perturbative CEM parameters.
In section \ref{sec:results}, we then present our results for the production of di-$J/\psi$  for the CMS, ATLAS, and LHCb
acceptances, of  $J/\psi+\Upsilon$ in the D0 acceptance and for the di-$\Upsilon$ in the CMS acceptance.
Section \ref{sec:conclusion} is devoted to our conclusions and outlook.

\section{The CEM in a few formulae\label{sec:NLOCEM}}

In the CEM, a given quarkonium-production cross section is obtained from that to produce the corresponding 
heavy quark-antiquark pair $Q \bar Q$ with the sole constraint that its invariant mass lies between twice the quark mass, $2m_Q$, and twice that of the lightest open-heavy-flavour hadron, $2m_{H}$. The same logic applies in the case of a pair of quarkonia.
The cross section for single quarkonium production is then given by 
\eqs{d\sigma^{\rm (N)LO}_{\cal Q}= \P^{\rm (N)LO}_{\cal Q}\int_{2m_Q}^{2m_H} 
\frac{d\sigma_{Q\bar Q}^{\rm (N)LO}}{d m_{Q\bar Q}}d m_{Q\bar Q},
\label{eq:sigma_CEM}}
and that for the production of a pair, $\Q_1+\Q_2$, of  quarkonia
\eqs{d\sigma^{\rm (N)LO}_{\Q_1+\Q_2}= \! 
\prod_{i=1}^2 \P^{\rm (N)LO}_{\Q_i}\!\int_{2m_{Q_i}}^{2m_{H_i}} \!\!d m_{Q_i\bar Q_i} 
\frac{d\sigma_{Q_1\bar Q_1+Q_2\bar Q_2}^{\rm (N)LO}}{d m_{Q_1\bar Q_1}d m_{Q_2\bar Q_2}},
\label{eq:sigma_double_CEM}}
where ${d\sigma_{Q\bar Q}^{\rm (N)LO}}/{d m_{Q\bar Q}}$ \big(${d\sigma_{Q_1\bar Q_1+Q_2\bar Q_2}^{\rm (N)LO}}/{(d m_{Q_1\bar Q_1}d m_{Q_2\bar Q_2}})$\big) is the corresponding (doubly) differential cross section for $Q\bar Q$ ($Q_1\bar Q_1+Q_2\bar Q_2$) production as a function of the pair invariant mass(es), $m_{Q\bar Q}$ ($m_{Q_1\bar Q_1}  m_{Q_2\bar Q_2}$) and $\P_{\Q_i}$ is a non-perturbative parameter encapsulating the probability for the 
hadronisation of the $Q \bar Q$ pair into the quarkonium $\Q_i$. It is supposed to be universal and independent
of the production of the pair.

In principle, having the heavy-quark cross section differential in the invariant mass, $d\sigma/dm_{Q\bar Q}$ is sufficient to obtain
the short-distance part of the CEM for single or associated production and correspondingly for 
pair production. The automated tool {\small \sc MadGraph5\_aMC@NLO} with specific cuts can
provide such cross sections up to NLO accuracy, also differential in other variables, like the rapidity
or the transverse momentum of the $Q\bar Q$ pair which translates\footnote{In an improved
version of the CEM~\cite{Ma:2016exq}, the quarkonium momentum is taken as that of the pair rescaled by the ratio
of the quarkonium mass over the pair invariant mass. In the case of the $J/\psi$, it 
slightly modifies the $p_T$ spectrum up to about 15 GeV.} into that of the quarkonium $\Q$. Such cross sections
should just then be multiplied by the non-perturbative parameter $\P_{\Q_i}$ which is usally tuned
to match the single-quarkonium production data.

\section{The $p_T$-differential cross section for single-quarkonium hadroproduction at NLO}\label{sec:fitCEM}

The existing CEM studies of quarkonium  production at RHIC, the Tevatron and the LHC rely on a 
hard-scattering matrix element at one loop for inclusive heavy-quark 
production, namely $\alpha_s^3$ (see~\cite{Lansberg:2019adr} for a recent review).  This is based on the  
well-known  multi-differential MNR computation~\cite{Mangano:1991jk} using the aforementioned 
invariant-mass cut. At this order, a heavy-quark pair with a non-zero $p_T$ (irrespectively of
the invariant mass of the pair) comes from real-emission graphs, where a final light parton
recoils against the $Q\bar Q$ pair. The virtual-emission contributions do not contribute away
from $p_{T,Q\bar Q}=0$. Such existing computations for $p_{T,Q\bar Q}\neq 0$ 
are effectively Born-order or tree-level computation from the partonic processes 
$gg [q \bar q] \to (Q \bar Q) g$ or $gq \to (Q \bar Q) q$, and thus not effectively at NLO accuracy.
As a case in point, the renormalisation-scale dependence of the resulting cross section 
is simply that of the third power of $\alpha_s(\mu_R)$.

Thanks to {\small \sc MadGraph5\_aMC@NLO}, we are able to provide complete  
NLO CEM hadroproduction results for $d\sigma/dp_{T,\Q}$ by computing $pp \to (Q \bar Q)_{\rm CEM}$ + 1 parton 
up to $\alpha_s^4$ where the subscript indicates that  the pair invariant 
mass is integrated as in \ce{eq:sigma_CEM}. A first $J/\psi$ study was presented along with our $J/\psi+Z$ CEM 
computation~\cite{Lansberg:2016rcx}. Here we go further and consider in addition
 the $\psi(2S)$ and $\Upsilon(nS)$ cases. We also discuss in more details the resulting
CEM parameter depending on whether it is fit at mid or large $p_T$ or on the $p_T$-integrated yields.

\begin{table}[htb!]
\begin{center}\footnotesize\renewcommand{\arraystretch}{1.2} 
\begin{tabular}{lcc} 
\hline \hline
 & $\P_{\cal Q}^{\rm LO}$ & $\P_{\cal Q}^{\rm NLO}$ \vspace{0.2em}\\
\hline
\multicolumn{3}{l}{Fits from $d\sigma/dp_T$: LO at ${\cal O}(\alpha_s^3)$  \& NLO at ${\cal O}(\alpha_s^4)$}\\
\hline
\multicolumn{3}{l}{ATLAS~\cite{Aad:2015duc}:  $\sqrt{s} = 8$ TeV,  $|y_{\psi}| < 0.5$, $p_T \in [8.5: 20]$ GeV}\\
\hline
$J/\psi$ & $0.015^{+0.013}_{-0.07}  $ & $0.009^{+0.004}_{-0.002} $ \\
$\psi$(2S) & $0.005^{+0.004}_{-0.002} $ & $0.003^{+0.001}_{-0.001} $ \\
\hline
\multicolumn{3}{l}{ATLAS~\cite{Aad:2015duc}:  $\sqrt{s} = 8$ TeV,  $|y_{\psi}| < 0.5$, $p_T \in [8.5: 110]$ GeV}\\
\hline
$J/\psi$ & $0.008^{+0.004}_{-0.003} $ & $0.006^{+0.003}_{-0.001} $ \\
$\psi$(2S) & $0.003^{+0.002}_{-0.01} $ & $0.002^{+0.003}_{-0.0005} $ \\
\hline
\multicolumn{3}{l}{CMS~\cite{Khachatryan:2015qpa} $\sqrt{s}  = 7$ TeV, $|y_{\Upsilon}| < 1.2$, $p_{T,\Upsilon} \in [10: 20]$ GeV}\\
\hline
$\Upsilon$(1S) & $0.04^{+0.03}_{-0.02}$ & $0.02^{+0.01}_{-0.005}$ \\
$\Upsilon$(2S) & $0.02^{+0.02}_{-0.01}$ & $0.01^{+0.02}_{-0.005}$ \\
$\Upsilon$(3S) & $0.01^{+0.01}_{-0.005}$ & $0.006^{+0.003}_{-0.001}$ \\
\hline
\multicolumn{3}{l}{CMS~\cite{Khachatryan:2015qpa} $\sqrt{s}  = 7$ TeV, $|y_{\Upsilon}| < 1.2$, $p_{T,\Upsilon} \in [10: 100]$ GeV)}\\
\hline
$\Upsilon$(1S) & $0.018^{+0.08}_{-0.06}$ & $0.012^{+0.02}_{-0.02}$ \\
$\Upsilon$(2S) & $0.013^{+0.06}_{-0.05}$ & $0.008^{+0.002}_{-0.001}$ \\
$\Upsilon$(3S) & $0.008^{+0.005}_{-0.003}$ & $0.005^{+0.002}_{-0.001}$ \\
\hline
\multicolumn{3}{l}{Fits from $\sigma$: LO at ${\cal O}(\alpha_s^2$)  \& NLO at ${\cal O}(\alpha_s^3)$}\\
\hline
\multicolumn{3}{l}{ALICE~\cite{Acharya:2019lkw}: $\sqrt{s} = 5.02$ TeV, $|y_{\psi}| < 0.9$, $p_{T,\psi}$ integrated} \\
\hline
$J/\psi$ & $0.015 \div 0.08$ & $0.004\div0.035$ \\
$\psi$(2S) & $0.003 \div 0.013$ & $0.0008\div 0.006$ \\
\hline
\multicolumn{3}{l}{CMS~\cite{Chatrchyan:2013yna} $\sqrt{s}  = 7$ TeV,  $|y_{\Upsilon}| < 2.4$, $p_{T,\Upsilon}$ integrated}\\
\hline
$\Upsilon$(1S) & $0.07^{+0.10}_{-0.04}$ & $0.03^{+0.03}_{-0.02}$ \\
$\Upsilon$(2S) & $0.02^{+0.03}_{-0.01}$ & $0.01^{+0.01}_{-0.005}$ \\
$\Upsilon$(3S) & $0.01^{+0.02}_{-0.005}$ & $0.005^{+0.003}_{-0.002}$ \\
\hline
\hline\hline
\end{tabular}
\caption{\label{table:FCEM} 
The coefficients $\P_{\cal Q}$ obtained by fitting the experimental data for several quarkonia.
}
\end{center}
\end{table}

As what regards the parameters of our computation, they remain very standard. We have 
used the PDF set NLO NNPDF 3.0 set~\cite{Ball:2012cx} with $\alpha_s(M_Z)=0.118$ 
provided by LHAPDF~\cite{Buckley:2014ana} from which we have derived the PDF uncertainty. The latter remains 
negligible compared to the factorisation- and renormalisation-scale uncertainties, 
which are evaluated by varying them independently in the 
interval $\frac{1}{2}\mu_0\le \mu_R,\mu_F \le 2\mu_0$, where 
$\mu_0$ is identified to the quarkonium transverse mass, $m_{T,\Q} = \sqrt{(2 m_Q)^2 + p_T^2}$.

Like in~\cite{Lansberg:2016rcx}, we use $m_c=1.27$~GeV for charmonium 
production in the CEM as suggested in~\cite{Nelson:2012bc}. It is important
to note that the quark mass enters the cross section both via $d\sigma/dm_{Q\bar Q}$ and 
via the integration range.  Results with $m_c=1.5$ GeV are 
sometimes slightly different. However, when the CEM is tuned to data, the mass dependence
is mostly absorbed in the change of $\P_{\Q_i}$ and the physics conclusion
always remains nearly identical. For the bottomonia, we have used $m_b= 4.7$ GeV.
For the upper bounds of integrations, $2m_H$, we have used  3.728 GeV for $c\bar c$ and 10.56 GeV
for $b\bar b$.

\begin{figure}[!htb]
\begin{center}
\subfloat[LO \& NLO $d\sigma/dp_T$ for $\psi(nS)$]{\includegraphics[width=0.95\columnwidth]{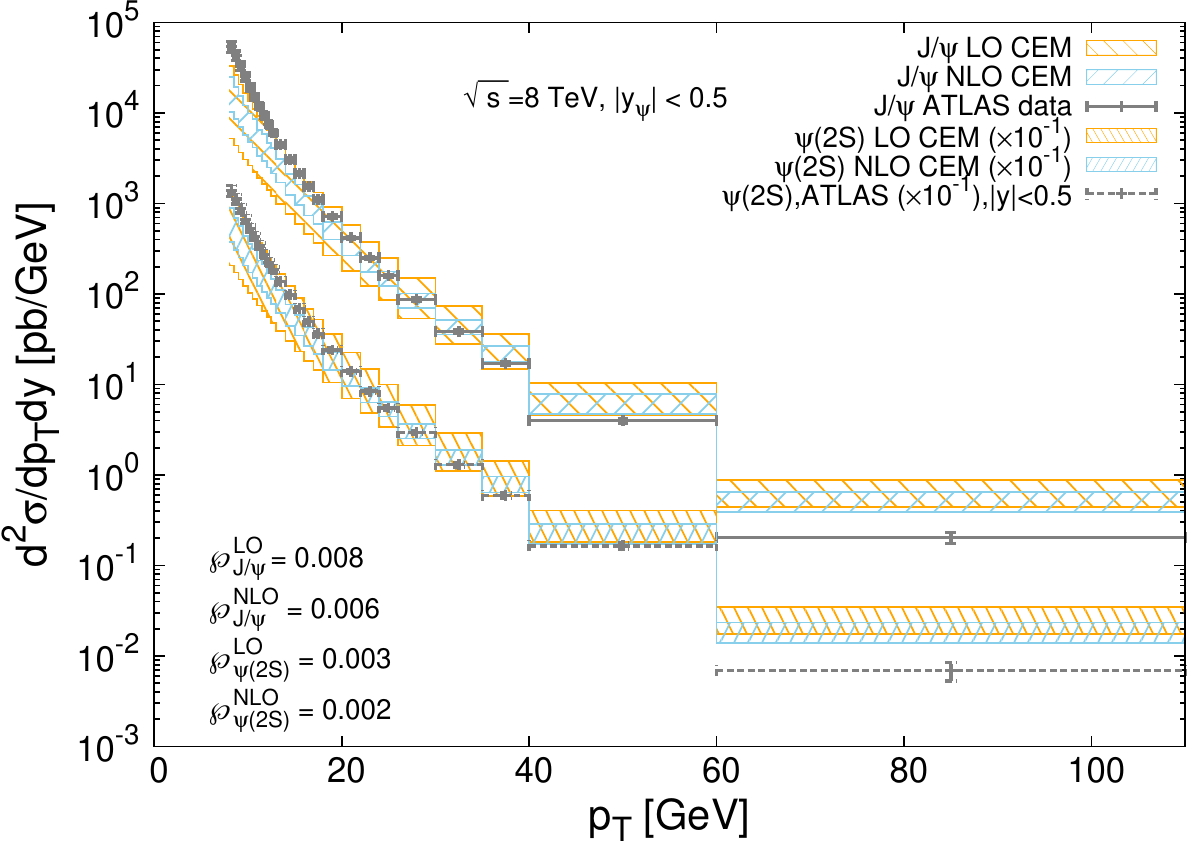}\label{fig:single_Jpsi_pT_ATLAS8TeV}}\\
\subfloat[LO \& NLO $d\sigma/dp_T$ for $\Upsilon(nS)$]{\includegraphics[width=0.95\columnwidth]{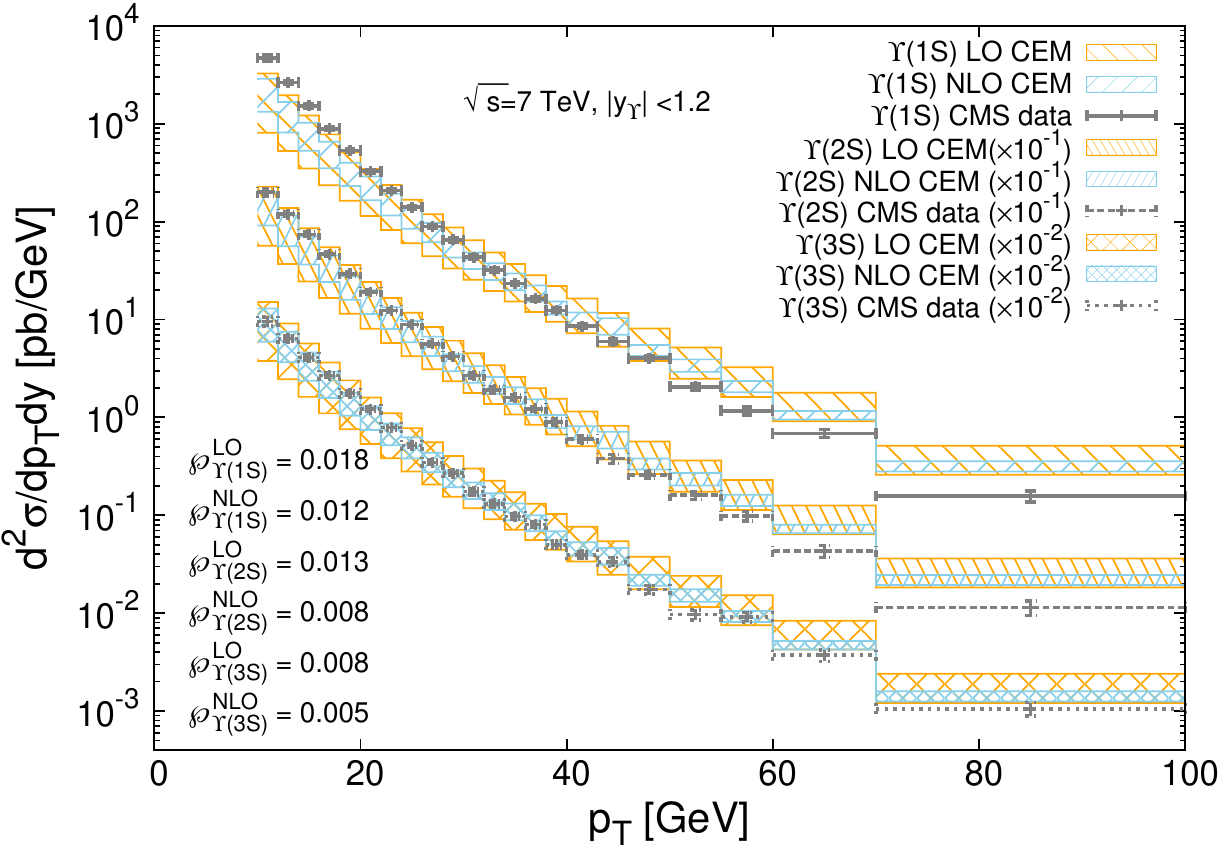}\label{fig:single_Upsilon_pT_CMS7TeV}}
\caption{
The $p_T$ (and $y$) differential cross sections of single (a) $J/\psi$ and $\psi$(2S) and
(b) $\Upsilon(nS)$ ($n=1,2,3$) production in the CEM.
The plotted data from ATLAS (8 TeV)~\cite{Aad:2015duc} and from CMS (7 TeV)~\cite{Khachatryan:2015qpa}
were used to fit $\P_{\cal Q}^{\rm (N)LO}$.
}
\end{center}
\end{figure}

We have performed a number of fits of $\P_{\cal Q}$ using the experimental data of single inclusive prompt $J/\psi$ and $\psi$(2S) and $\Upsilon(nS)$ data from ALICE, ATLAS and CMS over different $p_T$ ranges. We could also have used the very precise LHCb data~\cite{Aaij:2013yaa,Aaij:2015rla,Aaij:2019wfo}
but we preferred to restrict our fit to central rapidity data. Including them would not have changed our conclusions
since the CEM does not describe well the $p_T$ spectrum in any case.
 In the $J/\psi$ and $\Upsilon(nS)$ cases, the obtained values of $\P_{\Q_i}$ correspond to prompt production. For $\psi(2S)$, they hold for direct production. \ct{table:FCEM} gathers the used kinematical ranges and the corresponding fit results at LO and NLO.

Since
the $K$ factors for $pp\to c\bar c \text{ (+jet)} + X$ and $pp\to b \bar b \text{ (+jet)} + X$ near threshold are  larger than unity, the $\P_{\Q_i}$ at NLO are correspondingly smaller than at LO. Moreover, since the CEM $p_T$ spectra are usually too hard (see~\cite{Lansberg:2019adr}), the $\P_{\Q_i}$ also tend to decrease in order to match data at high $p_T$ and the fits overall worsen. This well-known (LO) trend is indeed confirmed at NLO. In the present LHC kinematics, this is particularly obvious in \cf{fig:single_Jpsi_pT_ATLAS8TeV} and \cf{fig:single_Upsilon_pT_CMS7TeV}. 

The ALICE $J/\psi$ data set and one CMS $\Upsilon$ data set extend to $p_T=0$ which allowed us to fit the $p_T$-integrated cross section with a NLO $\alpha_s^3$ computation of $pp\to (Q \bar Q)_{\rm CEM}+X$.
These results for the $\Upsilon$ case for the entire $p_T$ range are comparable to those of Vogt, 
\ie\ 2 to 3\% depending on $m_b$, the scale choice and the PDFs (see "$\Upsilon 1$" and "$\Upsilon 4$" of Table 8 of Ref. \cite{Bedjidian:2004gd}).  We see that the CEM parameters obtained by fitting the $p_T$ spectra are systematically smaller than those
obtained by fitting $p_T$-integrated yields. For the 
$J/\psi$ case, we have quoted a range. Indeed, as can be seen in~\cite{Nelson:2012bc}, $\sigma^{\rm NLO}(c\bar c)$ shows a very large uncertainty, which even tends to increase at large $\sqrt{s}$ ending up to be as large as one order of magnitude. The 
obtained lower values are systematically much smaller than the open-charm data. If we were to fit the ALICE $J/\psi$ data 
with $\sigma(c\bar c_{\rm CEM})$ computed with the scale values corresponding to these lower values, the discrepancy would be absorbed in $\P_\psi$
which would become anomalously large, even above unity in some cases. This would be unphysical. Since the open-charm data systematically lie between the central and upper NLO values, we thus only quote in \ct{table:FCEM} the corresponding range for $\P_{{\psi}}$. It is in fact in line with the values quoted in~\cite{Nelson:2012bc} for $m_c=1.27$~GeV but for different (fixed) scales
and PDFs. The data sets used for these older fits are also obviously different.

\section{LO and NLO CEM results for di-quarkonium hadroproduction}\label{sec:results}

In this section, we present our LO and NLO CEM results for all the existing LHC and Tevatron results~\cite{Aaboud:2016fzt,Khachatryan:2014iia,Aaij:2016bqq,Abazov:2014qba,Khachatryan:2016ydm,Sirunyan:2020txn}, but the D0 analysis~\cite{Abazov:2014qba} for which no normalised distributions were released and the early LHCb analysis at $\sqrt{s}=7$ TeV~\cite{Aaij:2011yc} which we consider to be superseded by their 13 TeV analysis. The corresponding kinematical conditions are summarised in \ct{table:phasespace}.

Like for the NLO single-quarkonium study presented above, we employ the NLO NNPDF 3.0 set~\cite{Ball:2012cx}. 
The dependence of the result on the renormalisation $\mu_R$ and factorisation $\mu_F$ scales is quantified by varying them independently in the interval $\frac{1}{2}\mu_0\le \mu_R,\mu_F \le 2\mu_0$ where $\mu_0$ depends on the considered system. For charmonium and bottomonium pairs, 
it is fixed to be $\sqrt{(4 m_Q)^2 + p_T^2}$ where $p_T$ is randomly selected from one of the pair members. For charmonium+bottomonium, 
it is the average of the transverse masses, $0.5 \times (m_{T1}+m_{T2})$. We also do not consider
uncertainties from the heavy-quark mass as they are mostly absorbed in the CEM parameters, $\P_{{\cal Q}_i}$. 
This is surely the case for
the invariant-mass-integration region. The remaining uncertainty from the value of the hard matrix element may differ, but in view of the data-theory disagreements which we discuss next, we consider this approximation to be reasonable.

\begin{table}[!htb]
\begin{center} \footnotesize \renewcommand{\arraystretch}{1.2} 
\begin{tabular}{l|l}
\hline\hline
Data set & Kinematical conditions \\
\hline
$J/\psi + J/\psi$
&
\hfil $\sqrt{s} = 7$ TeV ($pp$)\hfil 
\\
(CMS inclusive) \cite{Khachatryan:2014iia}
&
$\bullet$ $p_{T \psi} > 6.5$ GeV  when $|y_{\psi}| < 1.2$
\\
&
$\bullet$ 4.5 GeV $< p_{T \psi} < 6.5$ GeV  when 
\\
&
$1.43 \times (3.25-\frac{p_{T \psi}}{2})  < |y_{\psi}|< 4.45-\frac{p_{T \psi}}{2}$ 
\\
&
$\bullet$ $p_{T \psi} > 4.5$ GeV 
when $1.43 < |y_{\psi}| < 2.2$ 
\\
\hline
$J/\psi + J/\psi$ 
&
\hfil $\sqrt{s} = 8$ TeV ($pp$) \hfil
\\
(ATLAS fiducial) \cite{Aaboud:2016fzt}
&
$\bullet$ $p_{T \psi} < 8.5$ GeV,  $|y_{\psi}| < 2.1$
\\
&
$\bullet$ $p_{T\mu} < 4$ GeV, $\eta_\mu < 2.3$ 
\\
\hline
$J/\psi + J/\psi$
&
\hfil $\sqrt{s} = 13$ TeV ($pp$) \hfil
\\
(LHCb inclusive) \cite{Aaij:2016bqq}
&
$\bullet$ $p_{T \psi} < 10$ GeV
\\
&
$\bullet$ $2.0 < y_{\psi} < 4.5$
\\
\hline
$J/\psi + \Upsilon$(nS)
&
\hfil $\sqrt{s} = 1.96$ TeV ($p\bar p$)\hfil
\\
(D0 fiducial) \cite{Abazov:2015fbl}
&
$\bullet$ $p_{T\mu} > 2$ GeV
\\
&
$\bullet$ $|\eta_{\mu}| > 2.0$
\\
\hline
$\Upsilon +\Upsilon$
&
\hfil $\sqrt{s} = 8$ TeV ($pp$)\hfil
\\
(CMS inclusive) \cite{Khachatryan:2016ydm}
&
$\bullet$ $|y_{\Upsilon}| < 2.0$
\\\hline
$\Upsilon +\Upsilon$
&
\hfil $\sqrt{s} = 13$ TeV ($pp$) \hfil 
\\
(CMS inclusive) \cite{Sirunyan:2020txn}
&
$\bullet$ $|y_{\Upsilon}| < 2.0$
\\
\hline\hline
\end{tabular}
\caption{\label{table:phasespace}
Phase-space definition of the measured fiducial/inclusive production cross-section following the geometrical acceptance of each experiment.
The fiducial cuts, \ie\ those on the transverse momentum and pseudorapidity of muons generated by the decay of $J/\psi$ or $\Upsilon$, are given in terms of $p_{T \mu}$ and $\eta_{\mu}$, respectively.
}
\end{center}
\end{table}

\begin{figure*}[!htb]
\begin{center}
\subfloat[]{\includegraphics[width=0.33\textwidth]{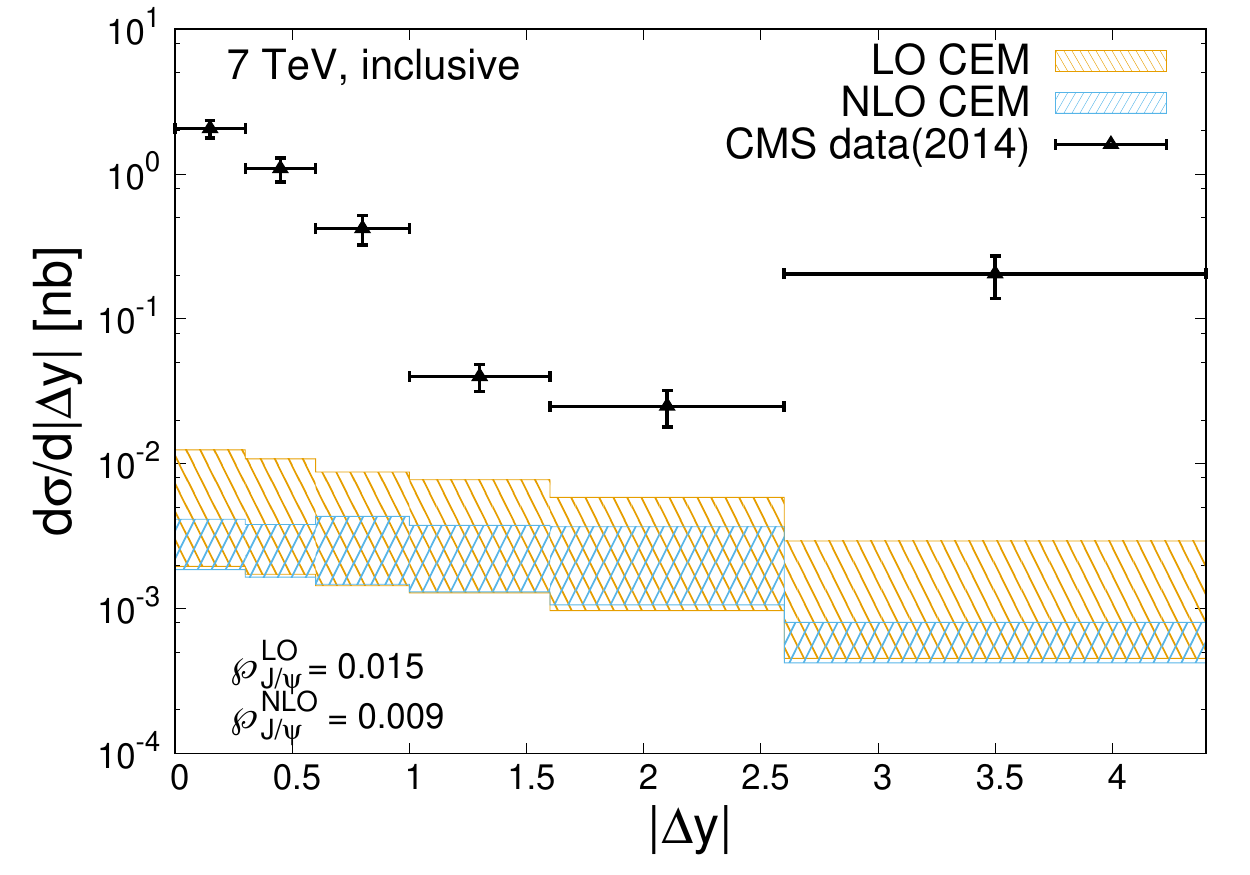}\label{fig:double_Jpsi_Deltay_CMS7TeV}}
\subfloat[]{\includegraphics[width=0.33\textwidth]{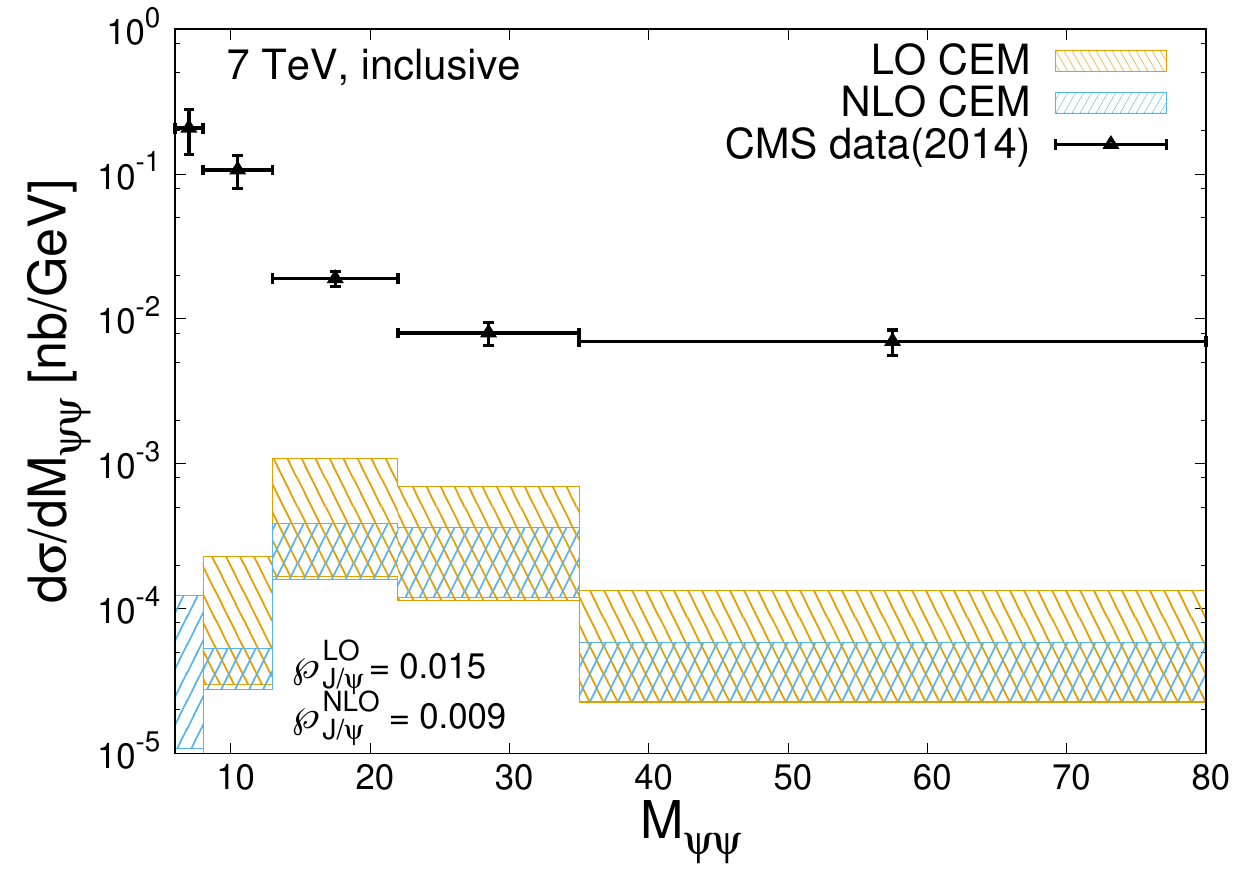}\label{fig:double_Jpsi_mpsipsi_CMS7TeV}}
\subfloat[]{\includegraphics[width=0.33\textwidth]{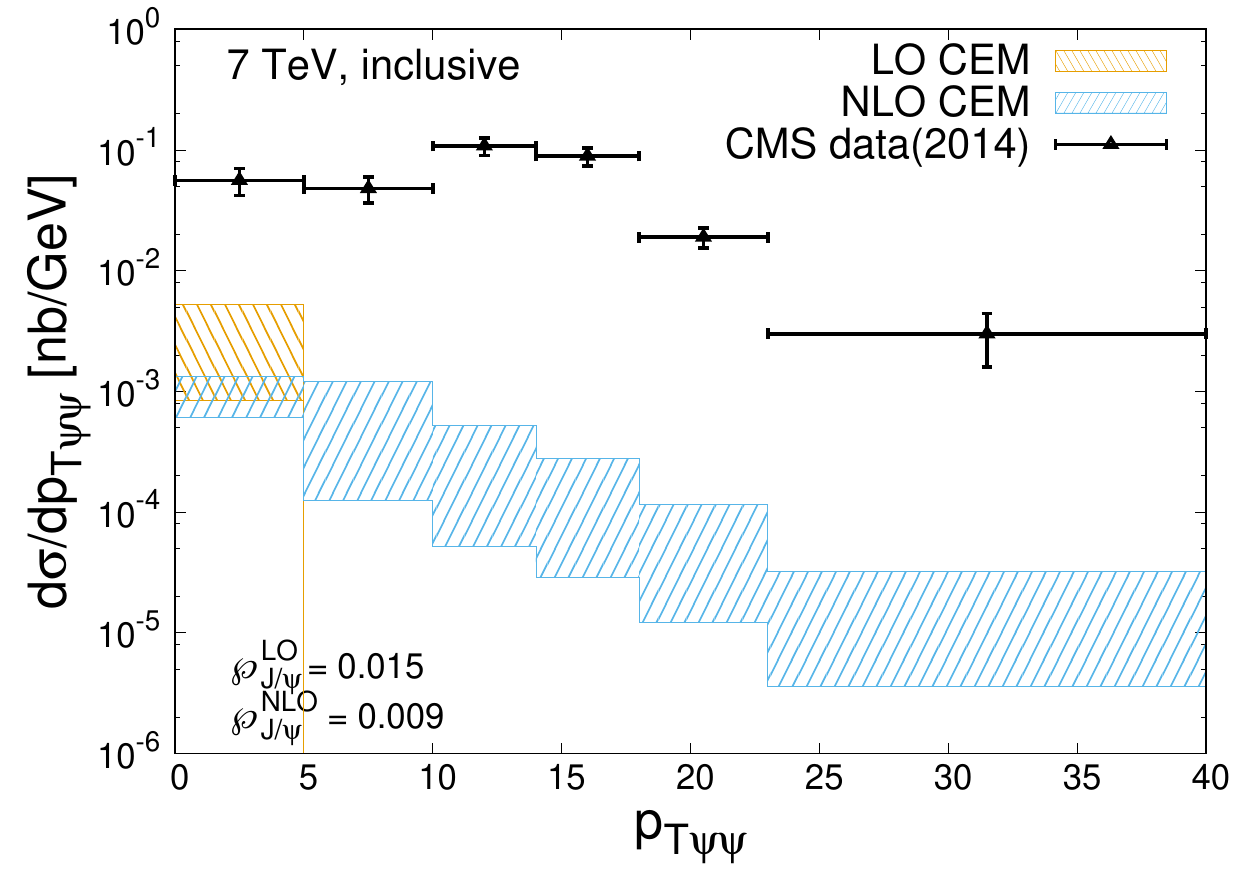}\label{fig:double_Jpsi_pTpsipsi_CMS7TeV}}\\
\subfloat[]{\includegraphics[width=0.33\textwidth]{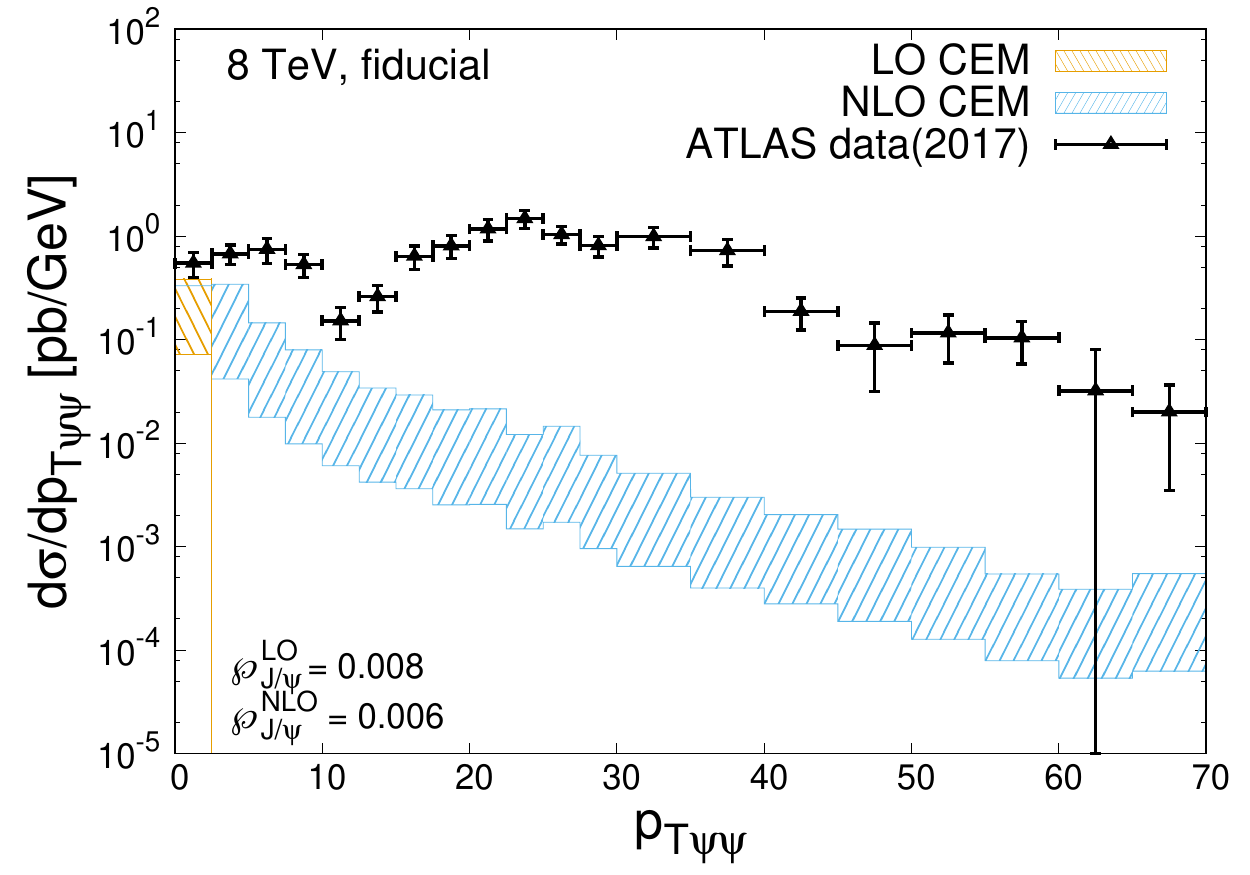}\label{fig:double_Jpsi_pTpsipsi_ATLAS8TeV-240420}}
\subfloat[]{\includegraphics[width=0.33\textwidth]{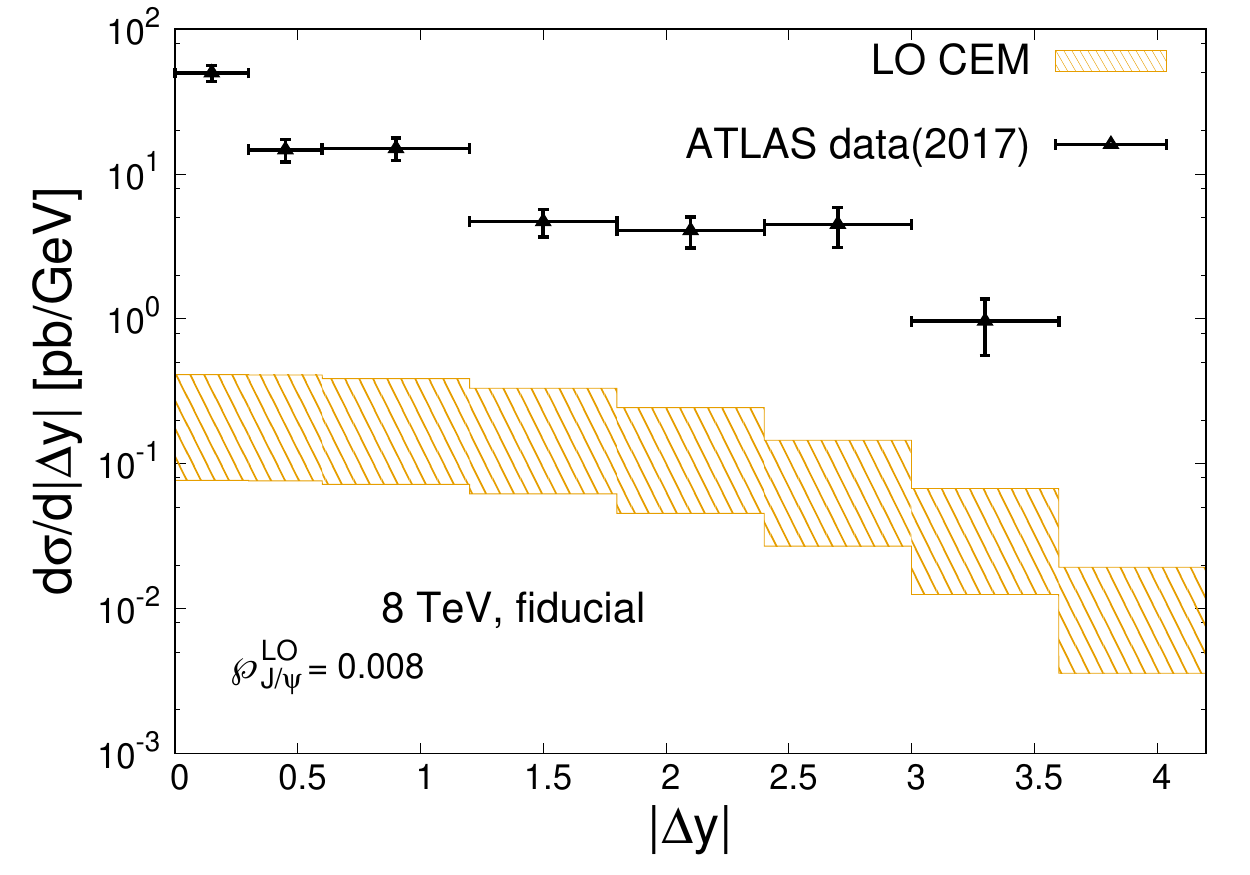}\label{fig:double_Jpsi_Deltay_ATLAS8TeV}}
\subfloat[]{\includegraphics[width=0.33\textwidth]{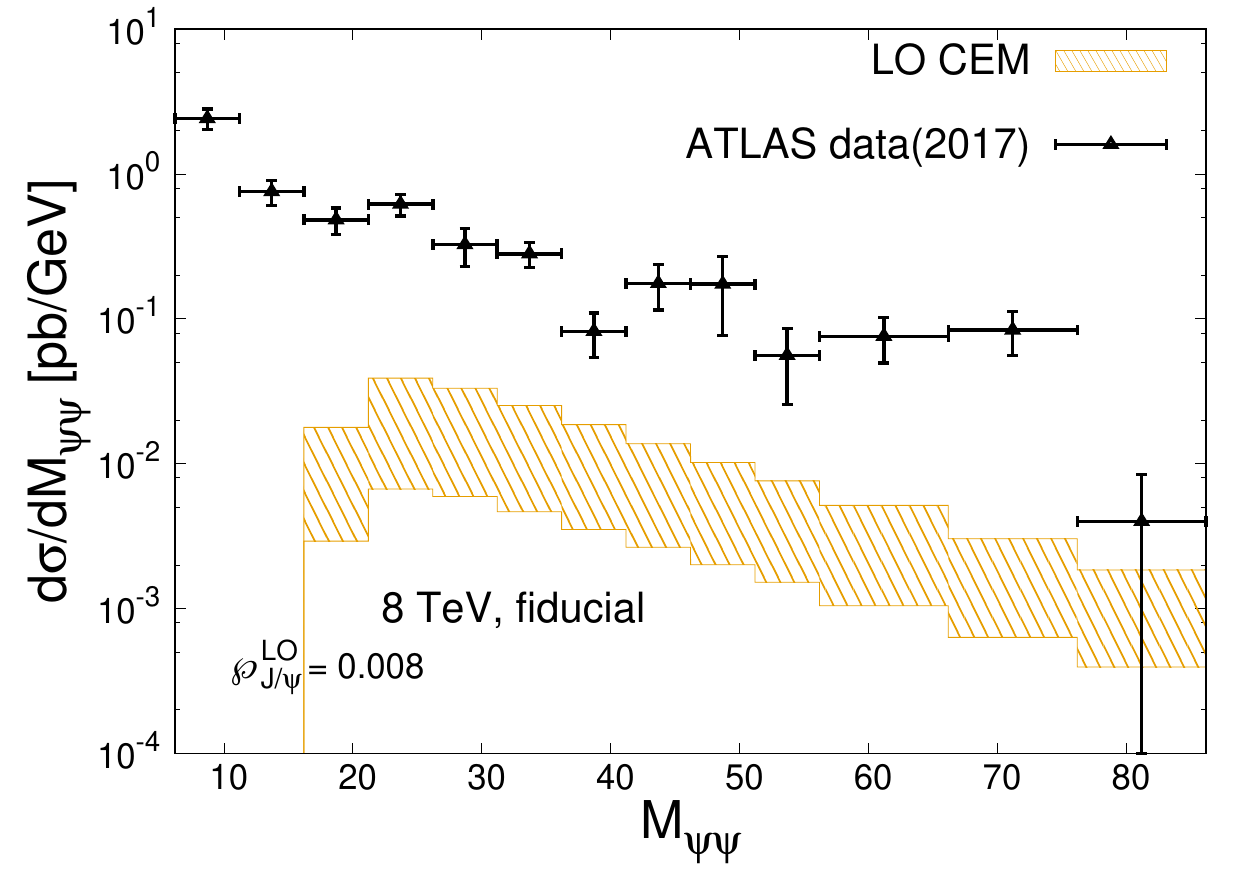}\label{fig:double_Jpsi_mpsipsi_ATLAS8TeV}}
\\
\subfloat[]{\includegraphics[width=0.33\textwidth]{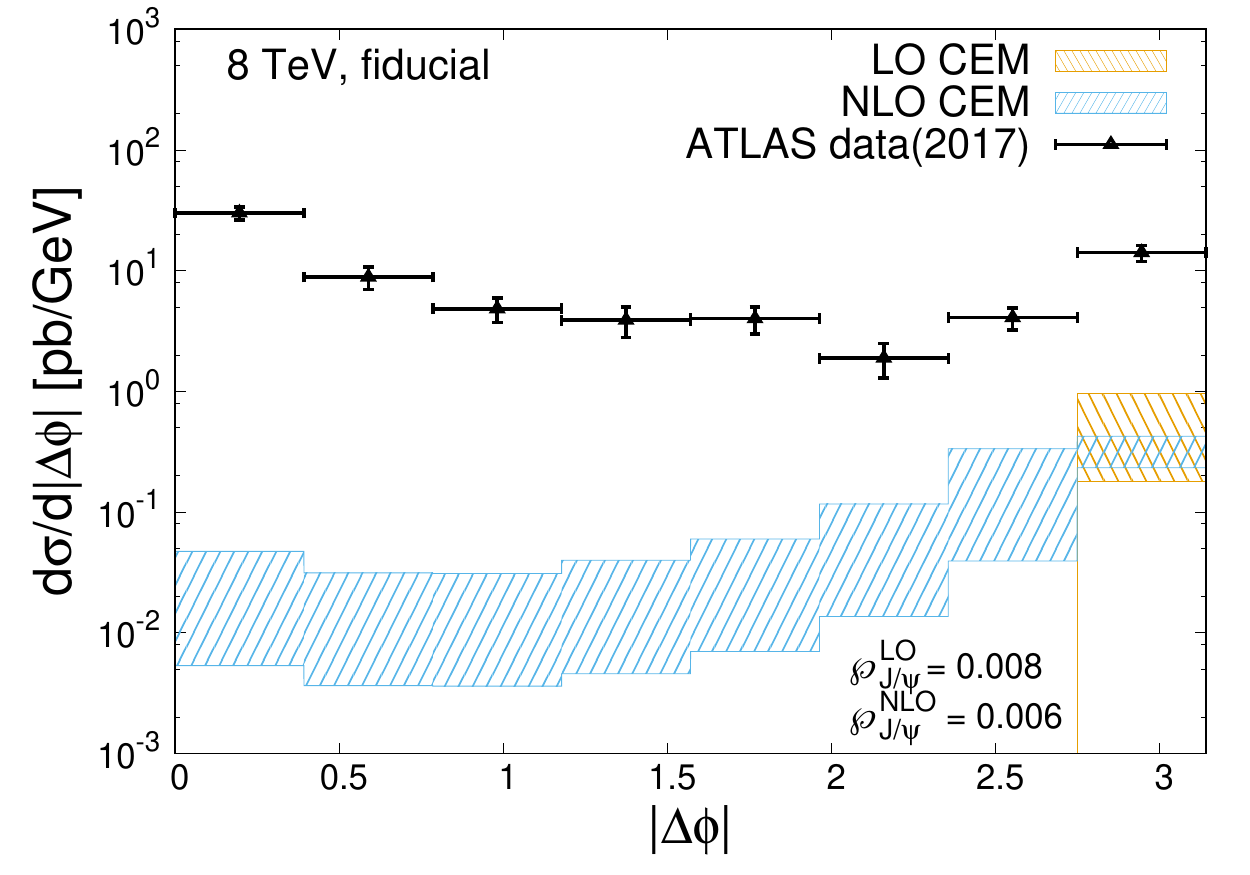}\label{fig:double_Jpsi_Deltaphi_ATLAS8TeV}}
\subfloat[]{\includegraphics[width=0.33\textwidth]{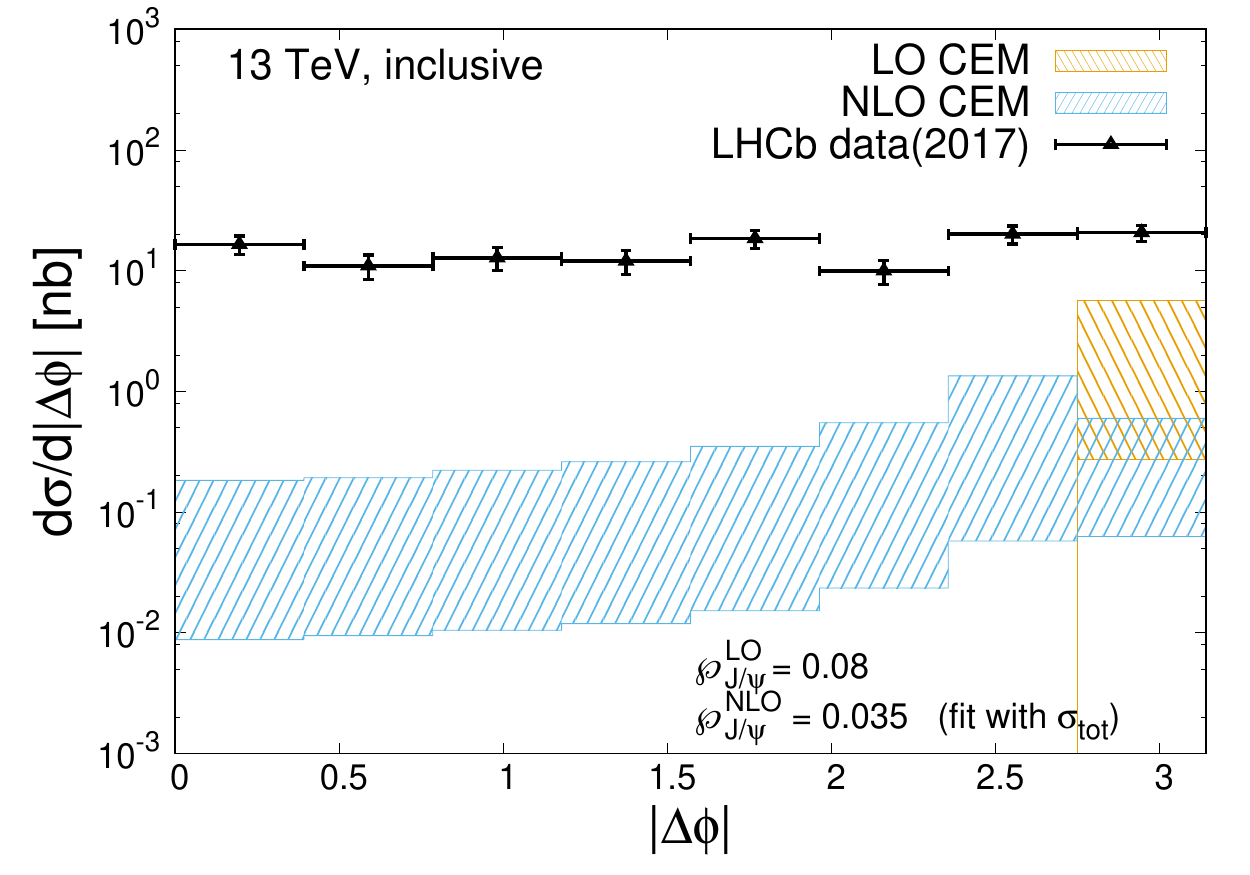}\label{fig:double_Jpsi_Deltaphi_LHCb13TeV}}
\subfloat[]{\includegraphics[width=0.33\textwidth]{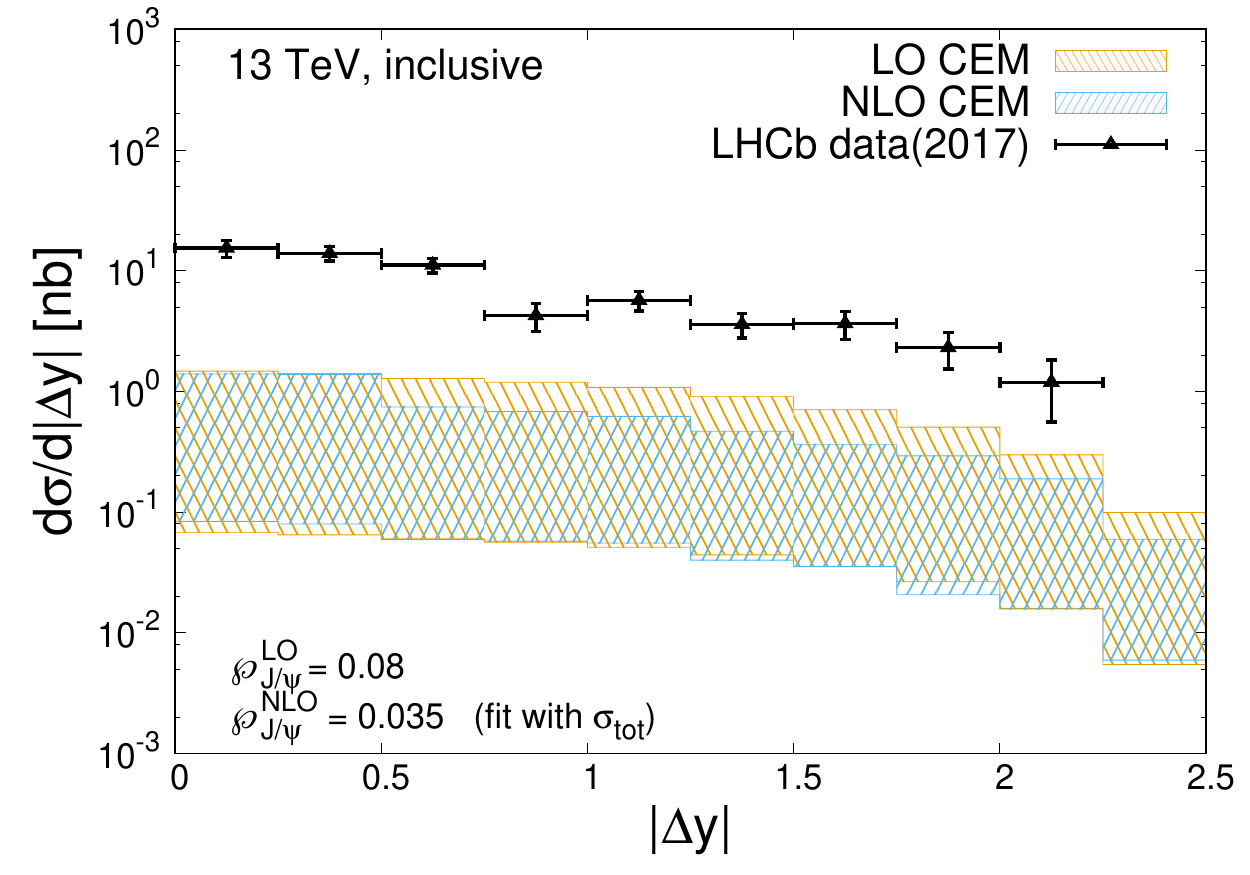}\label{fig:double_Jpsi_Deltay_LHCb13TeV}}\\
\subfloat[]{\includegraphics[width=0.33\textwidth]{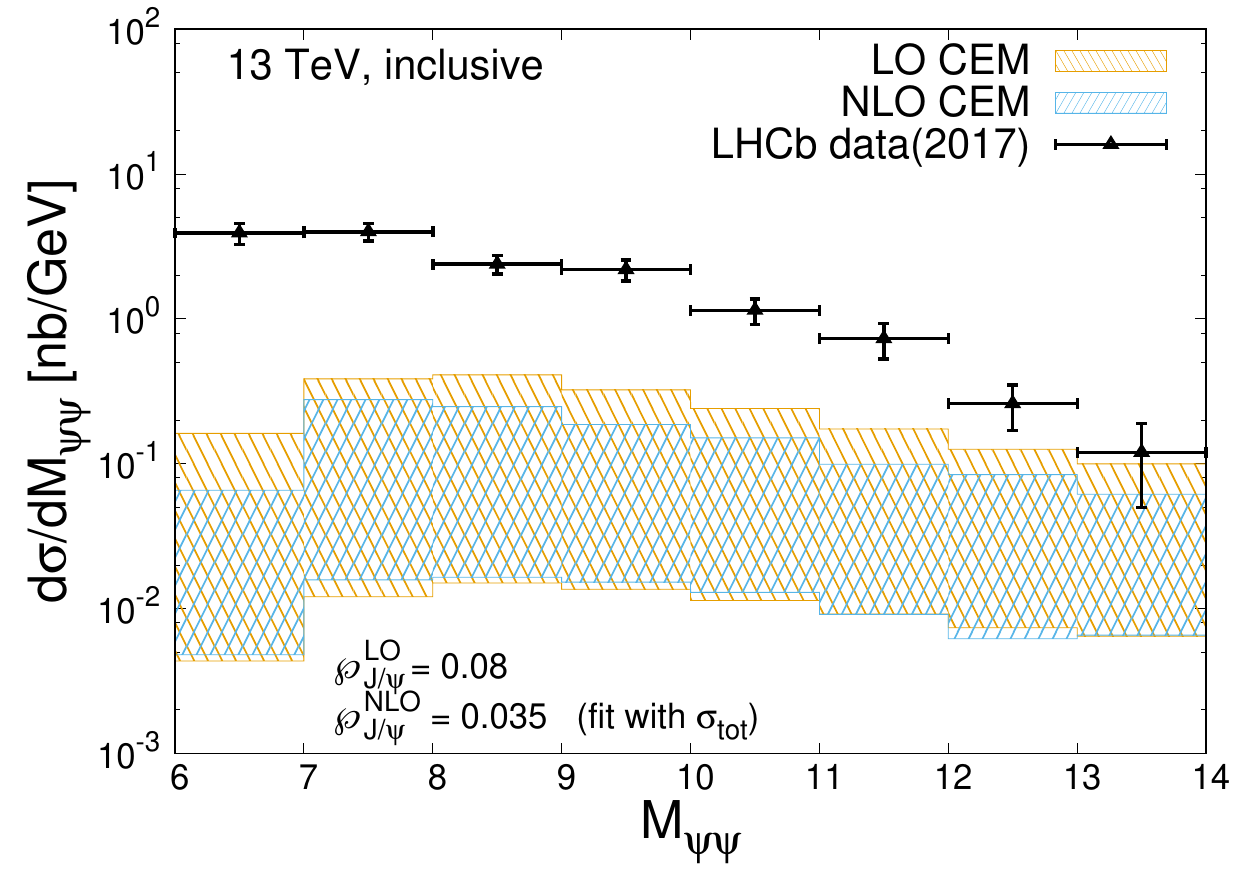}\label{fig:double_Jpsi_mpsipsi_LHCb13TeV}}
\subfloat[]{\includegraphics[width=0.33\textwidth]{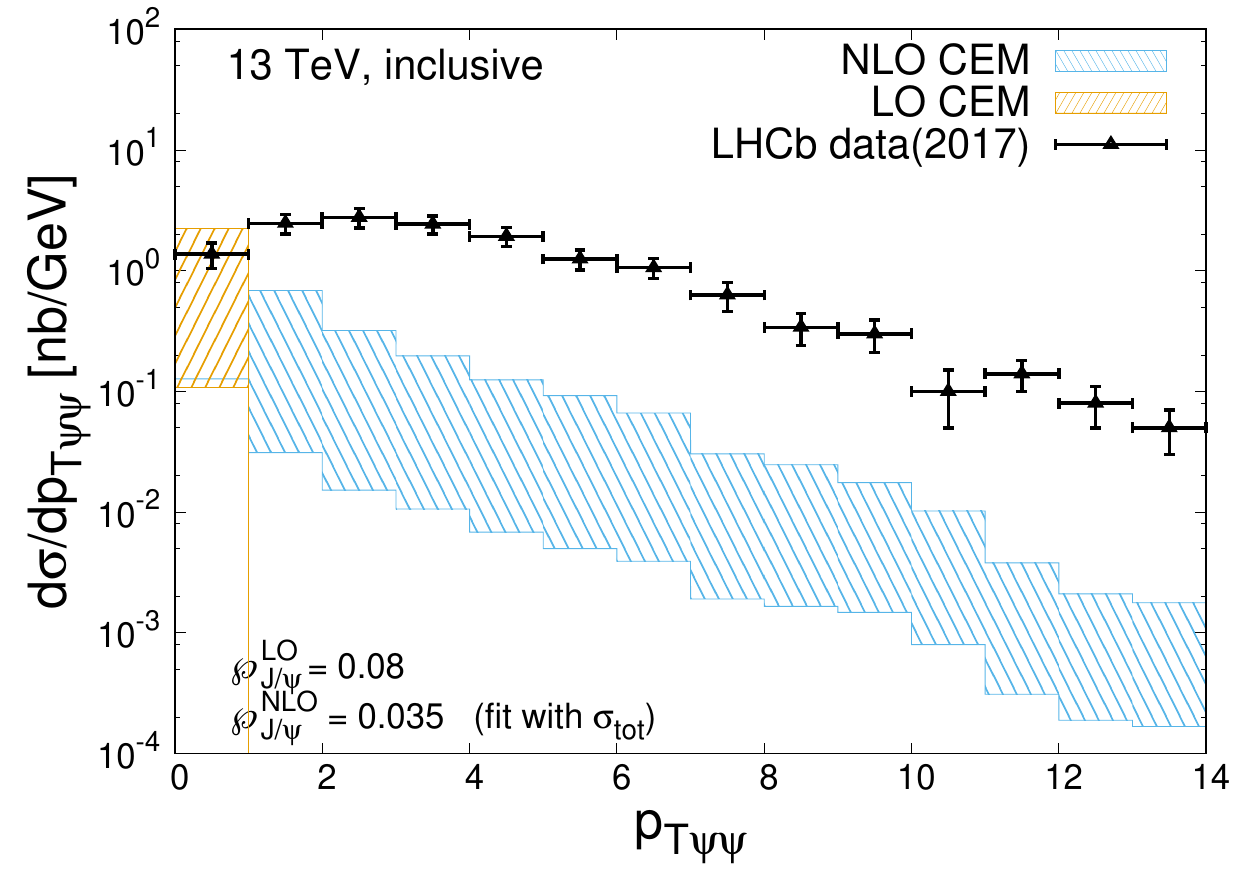}\label{fig:double_Jpsi_pTpsipsi_LHCb13TeV}}
\subfloat[]{\includegraphics[width=0.33\textwidth]{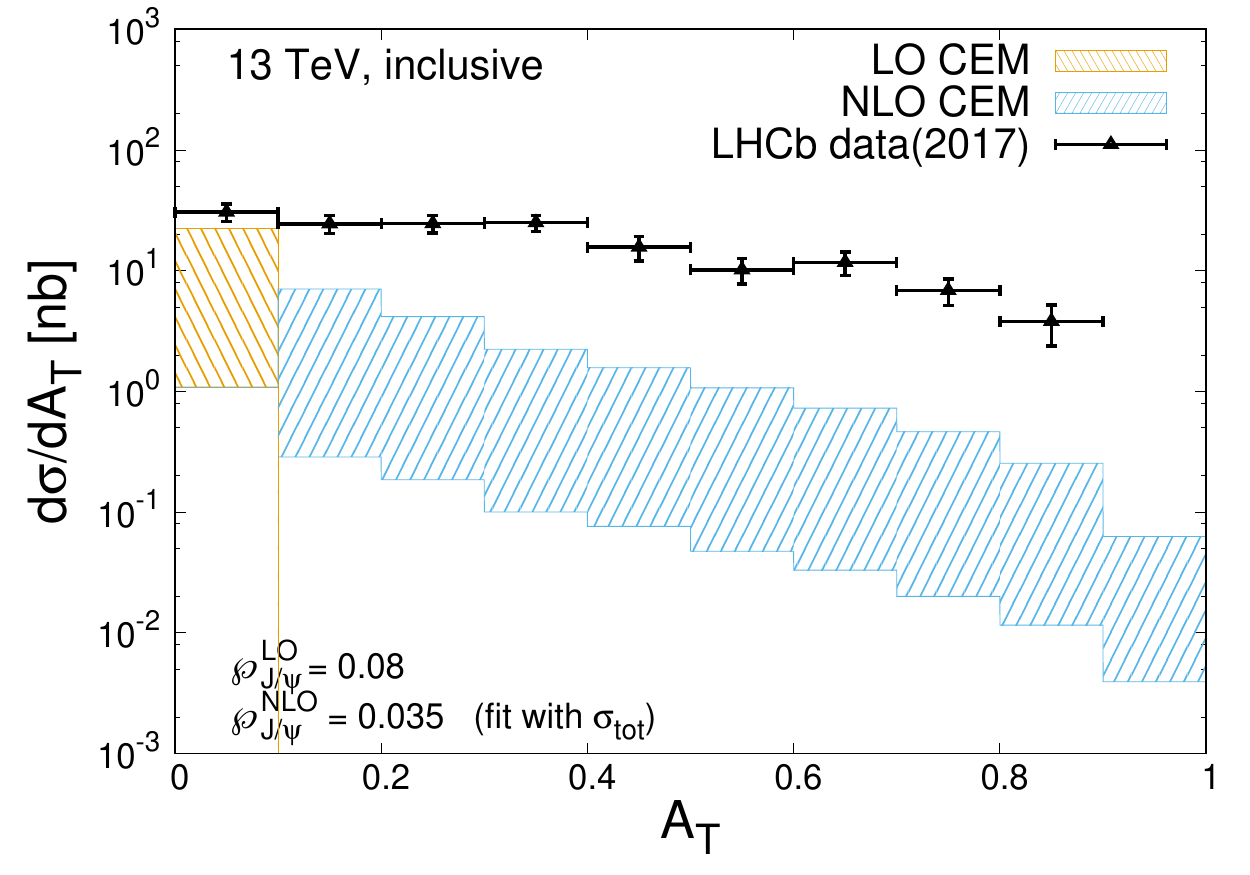}\label{fig:double_Jpsi_AT_LHCb13TeV}}
\caption{Various existing kinematical distributions of di-$J/\psi$ events compared to our LO and NLO CEM computations. See text for details.}
\label{fig:double_Jpsi}
\end{center}
\end{figure*}

\subsection{$J/\psi$ pairs}

Let us first discuss our results of the CEM calculation of $J/\psi$-pair production in the CMS setup.
The differential cross section in the rapidity separation, $|\Delta y_{{\psi \psi}}|$, is shown in \cf{fig:double_Jpsi_Deltay_CMS7TeV},
in the invariant mass, $M_{{\psi \psi}}$, in~\cf{fig:double_Jpsi_mpsipsi_CMS7TeV}, and in the transverse 
momentum of the $J/\psi$-pair, $p_{T{\psi \psi}}$, in \cf{fig:double_Jpsi_pTpsipsi_CMS7TeV}.
We see that the CEM results are at least an order of magnitude below the experimental data of CMS at both LO and NLO, even considering their (large) uncertainties. We note that the scale uncertainty in the NLO calculations is half of that in the LO ones which indicates the absence of kinematically-enhanced topologies. The regions of 
large $|\Delta y_{{\psi \psi}}|$ and large $M_{{\psi \psi}}$ are those where the DPS contributions are extracted and 
our computations confirm that SPS contributions, from the CEM for sure but likely as from other models, are negligible there.
Unsurprisingly, the NLO yield populates the $p_{T{\psi \psi}}$ distributions but its contribution is clearly too small and does not
even show the bump generated by the kinematical cut in the CMS acceptance. Such a bump is well seen in NLO CSM computations, which describes the data well all over the entire    spectrum~\cite{Lansberg:2014swa,Lansberg:2019fgm}.  The very same observations can be made for the $p_{T{\psi \psi}}$ distributions measured by ATLAS~\cite{Aaboud:2016fzt}.

\begin{figure*}[htb!]
\begin{center}
\subfloat[]{\includegraphics[width=0.66\columnwidth]{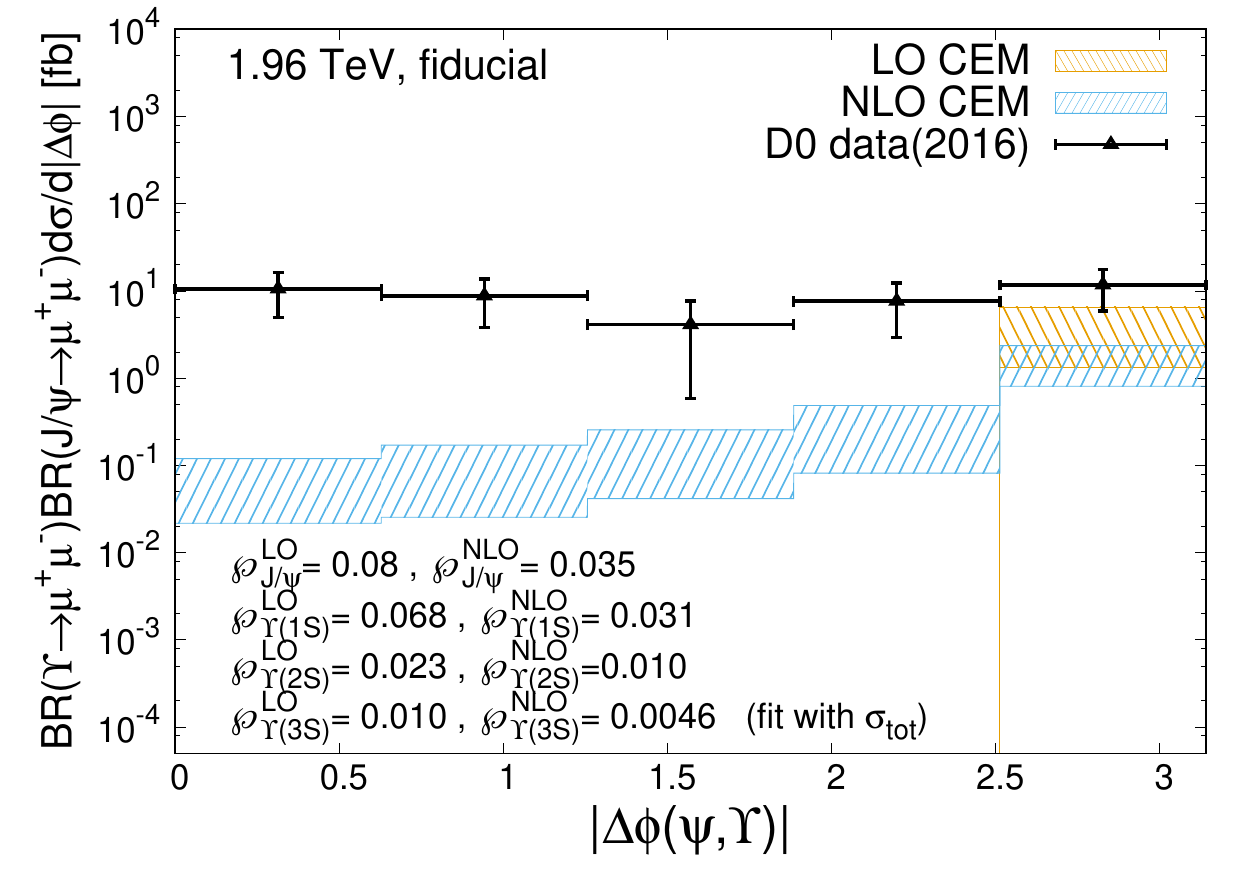}\label{fig:Jpsi_Upsilon_D0}}
\subfloat[]{\includegraphics[width=0.66\columnwidth]{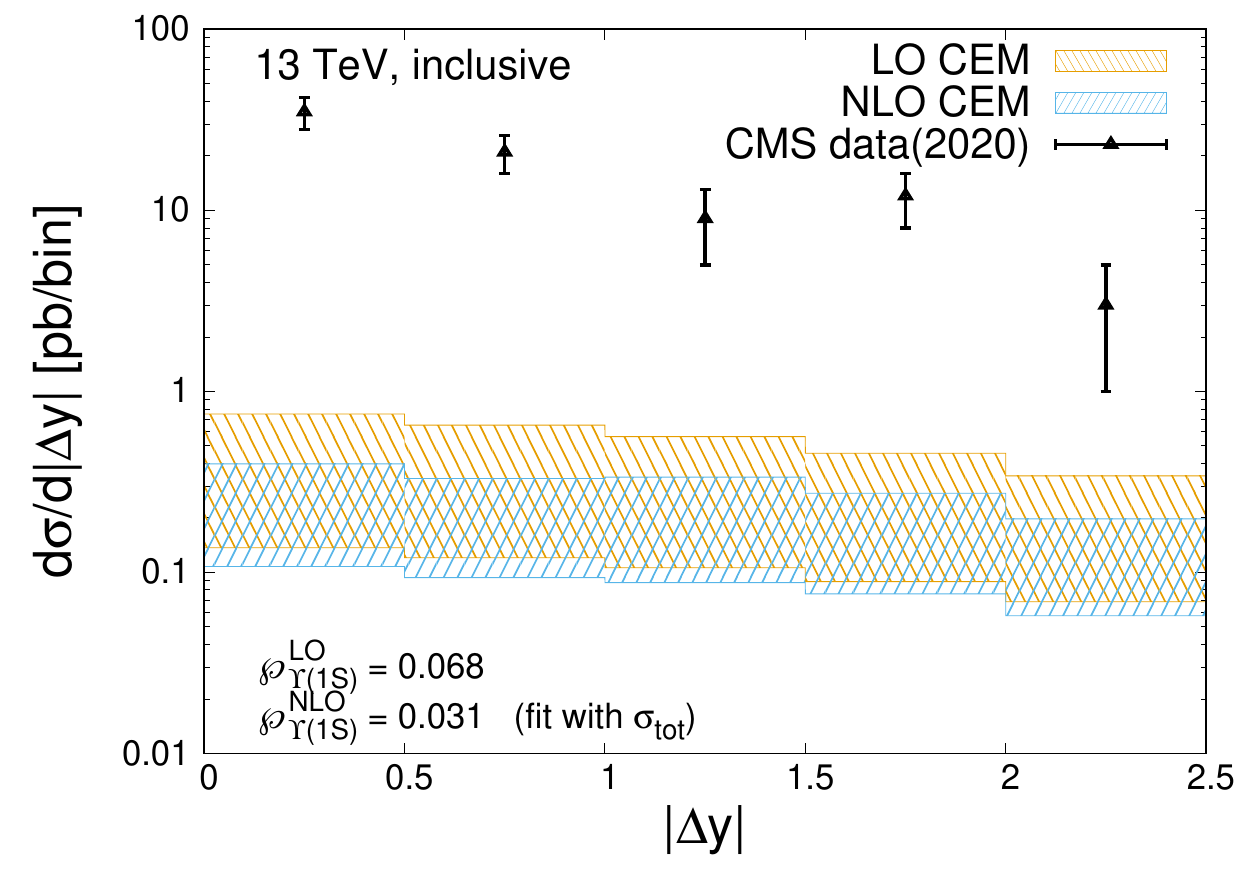}\label{fig:double_Upsilon_Deltay_CMS13TeV}}
\subfloat[]{\includegraphics[width=0.66\columnwidth]{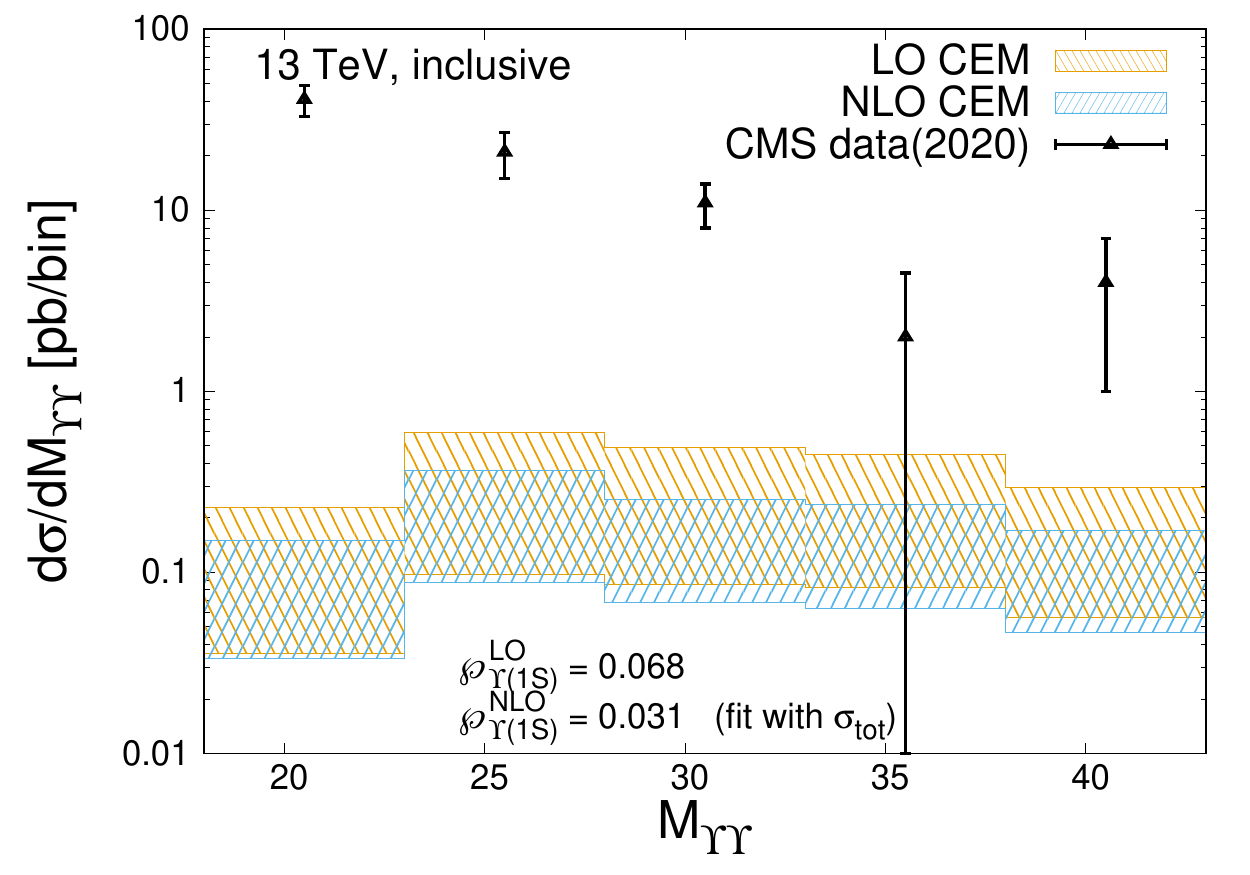}\label{fig:double_Upsilon_mupsiupsi_CMS13TeV}}
\caption{Comparison between our LO and NLO CEM computations and the experimental data for	
(a) the $|\Delta \phi|$ distribution of $J/\psi + \Upsilon$ at the Tevatron and (b) the $|\Delta y|$  \& (c) $M_{\Upsilon\Upsilon}$ distributions of di-$\Upsilon$ at the LHC.
}
\end{center}
\end{figure*}

We note that ATLAS only released the $M_{{\psi \psi}}$ and $|\Delta y_{{\psi \psi}}|$ distributions for their fiducial yields. For the latter, only a small fraction of the events passes the muon cuts, even after the $J/\psi$ cuts (which are also stringent, $p_{T{\psi}}>8.5$~GeV). In addition, the CEM yield is computed over a tiny fraction of the possible $c \bar c$ invariant masses. Overall, the relevant multi-dimensional hyperspace where the integration is performed can be extremely small and complex. The result depends on an extremely small part of the physical phase for $pp \to  c \bar c+ c \bar c+X$, especially at LHC energies. 
To help the integrator find the CEM `domain', one has to enlarge the invariant-mass regions at the MC generation level and then to restrict it at the histogramming level. This is unfortunately extremely ineffective. 
Whereas the {\small \sc MadGraph5\_aMC@NLO} integrator manages to perform well the integration at LO in a reasonable amount of time, it becomes highly CPU consuming at NLO, for instance ${\cal O}(10^8)$ CPU$\cdot$seconds to get distributions where all the bins are simply populated. Unfortunately, we did not manage to obtain reliable NLO results for these distributions. As such, we
only plot the LO results (see \cf{fig:double_Jpsi_Deltay_ATLAS8TeV}
for $|\Delta y_{{\psi \psi}}|$, \cf{fig:double_Jpsi_mpsipsi_ATLAS8TeV} for $M_{{\psi \psi}}$). We expect the NLO results to be similar in view of the LO/NLO ratio for the corresponding CMS distribution in
a similar (inclusive) phase space. 

Contrary to CMS, ATLAS also released their data as a function of $|\Delta \phi|$, the azimuthal angle between both $J/\psi$. It is
useful in quantifying the relative size of the DPS vs SPS contributions, especially when transverse-momentum-smearing effects 
can be neglected. Indeed, in such a case, the SPS contributions usually exhibit a peak at $|\Delta \phi|=\pi$ (both particles recoil on each other) and sometimes  at $|\Delta \phi|=0$ (the pair recoils against a third particle) whereas the DPS contributions 
should exhibit a flat distribution if both partonic scatterings are indeed uncorrelated. This remains of course 
an approximation although, until now, never falsified.

Along these lines, the concave data $|\Delta \phi|$ distribution shown in \cf{fig:double_Jpsi_Deltaphi_ATLAS8TeV} is indicative of a significant SPS contributions. According to ATLAS~\cite{Aaboud:2016fzt}, it amounts to about 90\% of the entire yield. Clearly, the CEM is unable to account for this SPS contribution. The same distribution measured by LHCb at 13 TeV shown on \cf{fig:double_Jpsi_Deltaphi_LHCb13TeV} is however much more intricate to interpret. Indeed, the LHCb measurement was performed without $p_{T\psi}$ cut which allows the momentum-smearing effects to be significant. As a consequence, the DPS vs SPS separation is much more involved. On a logarithmic plot, the NLO CEM yield already looks nearly flat, not very different than the shape of the data distribution. Yet, the normalisation is again off by more than an order of magnitude. Let us stress that, for the LHCb acceptance, we have used a $\P_{\cal Q}$ value fit on the $p_T$-integrated yields,
which is the largest of all those discussed above. One observes a similar gap on the other distributions\footnote{$A_T$, also called the transverse momentum asymmetry, is defined as $A_T=\left|\frac{p_{T\psi_1}-p_{T\psi_2}}{p_{T\psi_1}+p_{T\psi_2}}\right|$.} (see~\cf{fig:double_Jpsi_Deltay_LHCb13TeV}-\ref{fig:double_Jpsi_AT_LHCb13TeV}) which confirms that the CEM is unable to account for any measured di-$J/\psi$ data sets. This is
even the case in regions where at the same time the DPS contributions are expected to be mild  --if not negligible-- and the CSM
has been found to match the data.

\subsection{$\Upsilon+J/\psi$ pairs}

We now move to the $J/\psi + \Upsilon(nS)$ case as measured by the D0 Collaboration at $\sqrt{s}=1.96$~TeV~\cite{Abazov:2015fbl}. The only released kinematical distribution was that of $|\Delta \phi|$ which we have compared to our CEM computations in~\cf{fig:Jpsi_Upsilon_D0}. We note that this measurement was performed at low $p_T$ and we have used
the corresponding CEM parameter values. If we had used parameters fit to the $p_T$-differential data, the CEM predictions would have been even smaller. At $|\Delta \phi|=\pi$, the NLO CEM is at best 5 times below the data and ends up to be 100
 times lower at $|\Delta \phi|=0$. This is in line with the current interpretation of these D0 data, namely that they are dominated by
DPS contributions~\cite{Shao:2016wor}.

\subsection{$\Upsilon$ pairs}

Finally, we move to $\Upsilon$(1S)-pair production as measured by the CMS experiment.
In a first study at 8 TeV~\cite{Khachatryan:2016ydm} for $|y_\Upsilon|<2.0$, they only measured the integrated cross section, which they found to be 
\begin{equation}
\sigma^{\rm exp}_{\Upsilon \Upsilon}
=
[68.8 \pm 12.7 ({\rm stat}) \pm 7.4 ({\rm sys}) \pm 2.8 ({\rm Br})] \, {\rm pb}
,
\end{equation}
to be compared to our CEM results (with $\P_{\Upsilon(1S)}$ fit to the $p_T$-integrated spectrum, 0.068 and 0.031)
\begin{eqnarray}
\sigma^{\rm LO}_{\Upsilon \Upsilon}= 0.38^{+0.27}_{-0.17} \, {\rm pb} \text{~and~}
\sigma^{\rm NLO}_{\Upsilon \Upsilon}= 0.76^{+0.88}_{-0.41} \, {\rm pb}.
\end{eqnarray}

Very recently, CMS performed a new study at 13~TeV~\cite{Sirunyan:2020txn} with significantly more events which allowed them
to perform differential measurements as a function of $\Delta y$ and $M_{\Upsilon\Upsilon}$. The comparisons are shown in~\cf{fig:double_Upsilon_Deltay_CMS13TeV} \& \ref{fig:double_Upsilon_mupsiupsi_CMS13TeV}. 

As of now, there does not exist any direct or indirect DPS/SPS separation. As such, we are not able to claim
that the CEM is in contradiction with the data. Yet, any reasonable estimate of the DPS yield
would indicate that the SPS fraction should be significant~\cite{Lansberg:2019adr}. This would mean that
the CEM indeed cannot account for the SPS yield to di-$\Upsilon$ in the CMS acceptance.

\section{Conclusions\label{sec:conclusion}}

We have presented the first CEM study at LO and NLO for the SPS yields in hadroproduction of quarkonium pairs.
Our computation --fully accounting for contributions up to $\alpha_s^5$-- was performed thanks to a tuned version of
{\small \sc MadGraph5\_aMC@NLO} taking into account the specificities of the CEM. Except for those kinematical distributions
where the LO distributions are trivially suppressed, the $K$-factors we have found are systematically on the order of the unity, in
particular at large $\Delta y_{\Q\Q}$ and $M_{\Q\Q}$. This lends support to the irrelevance of possible kinematically-enhanced contributions from QCD corrections in these regions (see also~\cite{Baranov:2012re,Cisek:2017ikn,Babiarz:2019pth,Lansberg:2019fgm})  where the dominance of DPS contributions is sometimes questioned. Owing to the similarities between the CEM and the COM of NRQCD, we foresee a similar situation  when the first NLO COM studies are performed. 

On the quantitative level, we have compared our computations to a large selection of data for $J/\psi+J/\psi$, $J/\psi+\Upsilon$, and $\Upsilon+\Upsilon$ hadroproduction in $pp$ and $p \bar p$ collisions at the LHC and the Tevatron as measured by the ATLAS, CMS, D0, and LHCb collaborations. In all the cases, the computed yields are one to two orders of magnitude below the experimental data. This is also the case in the kinematical regions where it is established that the SPS contributions are dominant, or equivalently that the DPS contributions cannot reasonably describe the data. As such, it provides evidence that the CEM does not encapsulate the leading production mechanism for this SPS yield. This is another case, with $J/\psi+c\bar c$~\cite{Lansberg:2019adr}, where the CEM fails to describe the data while the CSM can.

In order to present coherent results at one-loop accuracy, we have also studied $p_T$-differential
cross sections for {\it single} quarkonium hadroproduction up to $\alpha_s^4$. We have used these in order to fit the non-perturbative CEM parameters. 
As far as the description of $p_T$-differential yields are concerned, our results therefore naturally supersede existing CEM results available in the literature (see \eg~\cite{Bedjidian:2004gd}) which
were only performed up to $\alpha_s^3$. We have also updated our $J/\psi$ NLO results made 
along a previous $J/\psi+Z$ NLO CEM study~\cite{Lansberg:2016rcx}. 
Let us add that this is the first time that $p_T$-differential
cross sections for $\Upsilon(nS)$ are computed at this order and compared to the data. 

Overall, the CEM features for {\it single} quarkonium hadroproduction observed at $\alpha_s^3$, \ie\ at LO, are confirmed and the CEM remains unable to provide a satisfactory
description of the single-quarkonium-hadroproduction data with too hard a spectrum at large $p_T$. For this reason, we provide different
values of the CEM parameters as needed makeshift if one wants to still perform other phenomenological studies
similar to the present one for di-quarkonium production. Indeed, the CEM still represents a handy approach to estimate
quarkonium cross sections when computations under other approaches like the CSM and NRQCD are too complex, 
especially beyond LO accuracy.


\section*{Acknowledgements} 
We thank C. Caillol, C. Flore, O. Mattelaer,  M.A. Ozcelik and  F. Scarpa for useful discussions.
This project has received funding from the European Union's Horizon 2020 research 
and innovation programme under grant agreement No 824093 in order to contribute to
the EU Virtual Access {\sc \small NLOAccess} and from the Franco-Chinese LIA FCPPL (Quarkonium4AFTER). 
The work of J.P.L. is supported in part by the French IN2P3--CNRS via the project GLUE@NLO. 
H.S.S. is supported by the ILP Labex (ANR-11-IDEX-0004-02, ANR-10-LABX-63).
N.Y. was supported by JSPS Postdoctoral Fellowships for Research Abroad.
Y.J.Z is supported by the National Natural Science Foundation of China (Grants No. 11722539).

\bibliographystyle{utphys}

\bibliography{double_Onium-CEM-260420}

\appendix

\end{document}